\documentclass[preprint2,times,tighten]{aastex6}
\usepackage{amsmath,amstext}
\usepackage[all]{hypcap} 

\shorttitle{PS1 RR Lyrae Stars}
\shortauthors{Sesar et al.}

\begin{document}

\title{Machine-Learned Identification of RR Lyrae Stars from Sparse, Multi-band Data: the PS1 Sample}

\author{Branimir Sesar\altaffilmark{1}}
\author{Nina Hernitschek\altaffilmark{1}}
\author{Sandra Mitrovi\'c\altaffilmark{1}}
\author{\v{Z}eljko Ivezi\'c\altaffilmark{2}}
\author{Hans-Walter Rix\altaffilmark{1}}
\author{Judith G.~Cohen\altaffilmark{3}}
\author{Edouard J.~Bernard\altaffilmark{4}}
\author{Eva K.~Grebel\altaffilmark{5}}
\author{Nicolas F.~Martin\altaffilmark{6,1}}
\author{Edward F.~Schlafly\altaffilmark{7,1}}
\author{William S.~Burgett\altaffilmark{8}}
\author{Peter W.~Draper\altaffilmark{9}}
\author{Heather Flewelling\altaffilmark{10}}
\author{Nick Kaiser\altaffilmark{10}}
\author{Rolf P.~Kudritzki\altaffilmark{10}}
\author{Eugene A.~Magnier\altaffilmark{10}}
\author{Nigel Metcalfe\altaffilmark{9}}
\author{John L.~Tonry\altaffilmark{10}}
\author{Christopher Waters\altaffilmark{10}}
\email{bsesar@mpia.de}

\altaffiltext{1}{Max Planck Institute for Astronomy, K\"{o}nigstuhl 17, D-69117 Heidelberg, Germany;}
\altaffiltext{2}{University of Washington, Department of Astronomy, P.O.~Box 351580, Seattle, WA 98195-1580}
\altaffiltext{3}{Division of Physics, Mathematics and Astronomy, Caltech, Pasadena, CA 91125}
\altaffiltext{4}{Universit\'{e} C\^{o}te d'Azur, OCA, CNRS, Lagrange, France}
\altaffiltext{5}{Astronomisches Rechen-Institut, Zentrum f\"ur Astronomie der Universit\"at Heidelberg, M\"onchhofstr.\ 12--14, 69120 Heidelberg, Germany}
\altaffiltext{6}{Observatoire astronomique de Strasbourg, Universit\'e de Strasbourg, CNRS, UMR 7550, 11 rue de l'Universit\'e, F-67000 Strasbourg, France}
\altaffiltext{7}{Lawrence Berkeley National Laboratory, One Cyclotron Road, Berkeley, CA 94720, USA}
\altaffiltext{8}{GMTO Corporation, 251 S.~Lake Ave., Suite 300, Pasadena, CA 91101, USA}
\altaffiltext{9}{Department of Physics, University of Durham, South Road, Durham DH1 3LE, UK}
\altaffiltext{10}{Institute for Astronomy, University of Hawai'i at Manoa, Honolulu, HI 96822, USA}

\begin{abstract}
RR Lyrae stars may be the best practical tracers of Galactic halo (sub-)structure and kinematics. The PanSTARRS1 (PS1) $3\pi$ survey offers multi-band, multi-epoch, precise photometry across much of the sky, but a robust identification of RR Lyrae stars in this data set poses a challenge, given PS1's sparse, asynchronous multi-band light curves ($\lesssim 12$ epochs in each of five bands, taken over a 4.5-year period). We present a novel template fitting technique that uses well-defined and physically motivated multi-band light curves of RR Lyrae stars, and demonstrate that we get accurate period estimates, precise to 2~sec in $>80\%$ of cases. We augment these light curve fits with other {\em features} from photometric time-series and provide them to progressively more detailed machine-learned classification models. From these models we are able to select the widest ($3/4$ of the sky) and deepest (reaching 120 kpc) sample of RR Lyrae stars to date. The PS1 sample of $\sim 45,000$ RRab stars is pure (90\%), and complete (80\% at 80 kpc) at high galactic latitudes. It also provides distances precise to 3\%, measured with newly derived period-luminosity relations for optical/near-infrared PS1 bands. With the addition of proper motions from {\em Gaia} and radial velocity measurements from multi-object spectroscopic surveys, we expect the PS1 sample of RR Lyrae stars to become the premier source for studying the structure, kinematics, and the gravitational potential of the Galactic halo. The techniques presented in this study should translate well to other sparse, multi-band data sets, such as those produced by the Dark Energy Survey and the upcoming Large Synoptic Survey Telescope Galactic plane sub-survey.
\end{abstract}

\keywords{methods: data analysis, statistical --- catalogs --- surveys --- stars: variables: RR Lyrae --- Galaxy: halo}
\maketitle

\section{Introduction}

Studies of the Galactic halo can help constrain the formation history of the Milky Way and the galaxy formation process in general. For example, in a recent theoretical study, \citet{har14} suggested there may be between 300 to 700 low luminosity ($<10^3$ $L_\sun$) dwarf satellite galaxies orbiting the Milky Way within 300 kpc of the Sun. However, the census of such low luminosity galaxies is currently complete only within $\sim45$ kpc of the Sun (Table 3 of \citealt{kop08}), and only at high galactic latitudes ($|b|>25\arcdeg$). A deeper, wider, and more complete census of Milky Way dwarfs would be extremely valuable, as it would allow us to test our assumptions about $\Lambda$CDM cosmology and galaxy formation, by comparing the observed distribution and properties of discovered dwarfs against those present in state-of-the-art hydrodynamic simulations, such as the APOSTLE \citep{saw16} and FIRE/\texttt{Gizmo} simulation suites \citep{hop14}.

The Galactic halo also contains remnants of accreted satellites (i.e., dwarf galaxies and globular clusters) that were disrupted by tidal forces and stretched into stellar tidal streams and clouds \citep[e.g., ][]{iba01, bel07, ses15, ber16}. Stellar streams are especially interesting because their orbits are sensitive to the properties of the Galactic potential (e.g., its shape and total mass of the Milky Way) and thus can be used to constrain it over the range of distances spanned by the streams (e.g., \citealt{kop10, new10, ses13, bel14}). For example, the total mass of the Milky Way is currently uncertain at a factor of two, and its more precise measurement using stellar streams may help resolve (or further aggravate) some apparent issues in the theory of galaxy formation, such as the so-called ``Too-Big-To-Fail'' problem \citep{bk11, wan12}. A more precise measurement of the total mass requires detailed modeling of stellar streams, such as the Sagittarius stream, as well as precise 3D kinematics and positions of stars in distant streams \citep{apw14}. While the Gaia mission \citep{per01} can and will deliver precise proper motions of halo stars, in most cases Gaia's parallax estimates will only be marginal beyond a few kiloparsec. Studies focusing on the 6D phase space structure of the Galactic halo will need to rely on tracers with precise (spectro-)photometric distances for the foreseeable future, making ``standard candles'' such as RR Lyrae stars enormously valuable.

To measure the total mass of the Milky Way and find the faintest dwarf satellites, we need to trace the spatial and kinematic structure and substructure (i.e., stellar streams) of the Galactic halo over the greatest possible distances and with the highest possible precision in distance and velocity, and the best tracers\footnote{Tracers are objects whose distribution reflects the distribution of the majority of stars (hopefully, in the least biased way).} for this task are RR Lyrae stars.

RR Lyrae stars are old (${\rm age}>10$ Gyr), metal-poor (${\rm [Fe/H] < -0.5}$ dex), pulsating horizontal branch stars with periodically variable light curves (periods ranging from 0.2 to 0.9 days; \citealt{smi04}). They are bright stars ($M_{\rm V}=0.6\pm0.1$ mag) with distinct light curves which makes them easy to identify with time-domain imaging surveys, even to large distances (5-120 kpc for surveys with a $14 < V < 21$ magnitude range; e.g., \citealt{ses10}). These properties, and the fact that {\em almost every Milky Way dwarf satellite galaxy has at least one RR Lyrae star} (including the faintest one, Segue 1; \citealt{sim11}), open up the exciting possibility of locating very low-luminosity Milky Way dwarf satellites by using distant RR Lyrae stars, as first proposed by \citet{ses14} (also see \citealt{bw15}).

RR Lyrae stars are also precise standard candles (i.e., their intrinsic brightness is well-determined). While distances to RR Lyrae stars can be measured with 3\% uncertainty using optical data (\autoref{distance_precision}), thanks to a tight period-luminosity relation in the near-infrared, distances to RR Lyrae stars can be measured with $2\%$ or better precision using, for example, $K$-band observations \citep{bra15, bea16}. Having precise distances is crucial for measuring tangential velocities\footnote{Radial velocities of RR Lyrae stars are straightforward to measure \citep{sesar12}.} and thus the Galactic potential, as the uncertainty in tangential velocity increases proportionally with the uncertainty in distance.

As made evident by several existing catalogs of Milky Way halo RR Lyrae stars (e.g., \citealt{viv01, mic08, kel08, ses10, ses13, dra13, aba14}), selection of RR Lyrae stars has become an almost routine procedure, as long as one has access to $\sim40$ or more observations per star {\em in a single photometric bandpass}. While very useful for many Galactic studies, the above catalogs are not ideal: they are either deep with limited sky coverage (e.g., the SDSS Stripe 82 catalog covers 100 deg$^2$ and is complete up to 110 kpc, \citealt{ses10}), or have wide coverage but are not very deep (e.g., the CRTS catalog covers 20,000 deg$^2$ and is complete up to 30 kpc, \citealt{dra13}). In addition, none of the above catalogs cover the Galactic plane, and thus cannot support studies of the old ($>10$ Gyr) Galactic disk. Currently the only existing imaging survey that has the potential to overcome all of the above drawbacks, and provide a deep and wide-area catalog of RR Lyrae stars in the northern skies (${\rm Dec}>-30\arcdeg$), is the Pan-STARRS1 (PS1) $3\pi$ survey.

\begin{figure}
\plotone{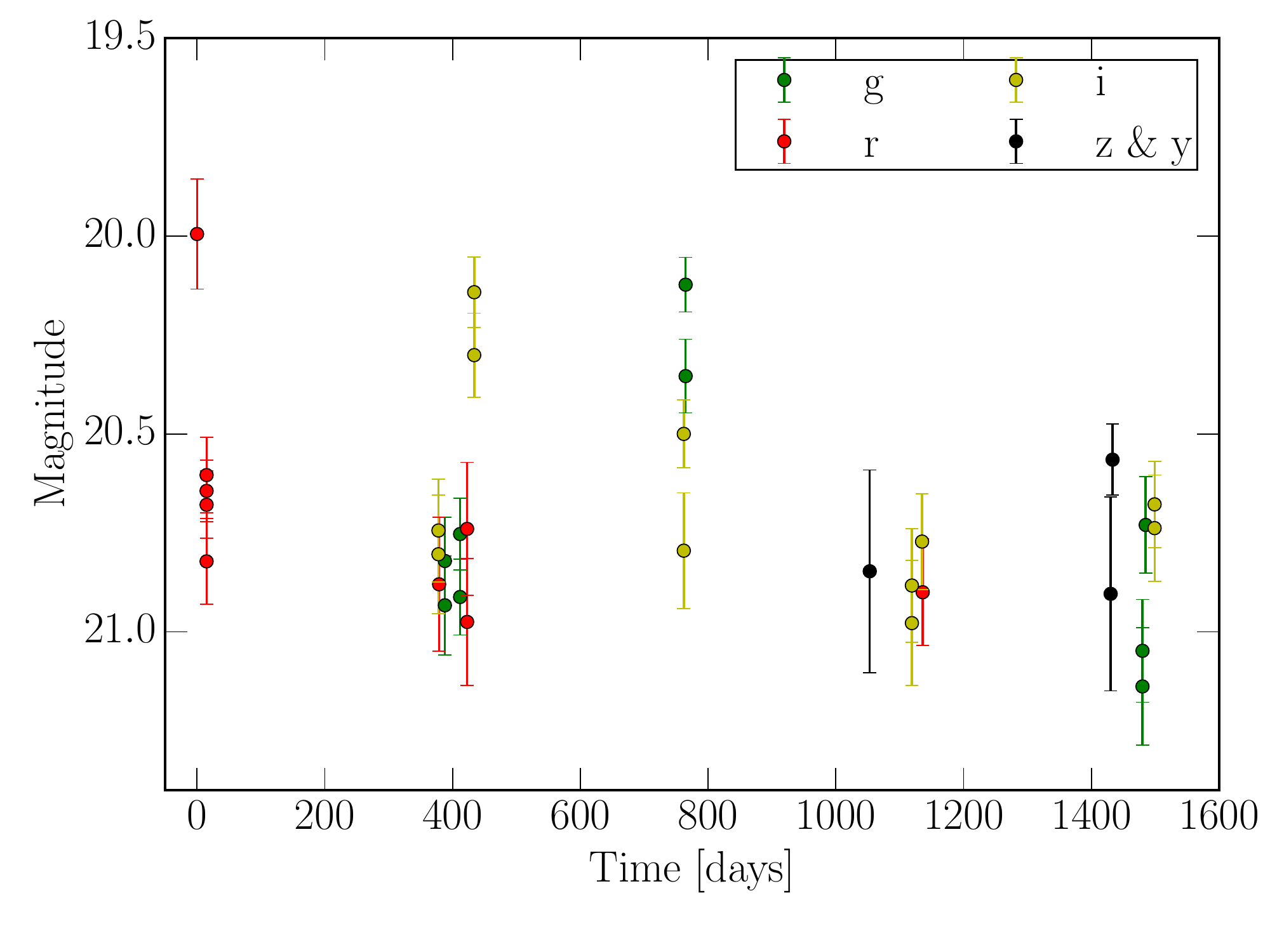}
\caption{
Multi-epoch PS1 $grizy$ photometry (i.e., light curves) of a faint RR Lyrae star. Note that the observations in different bands are not synchronous, and that the light curves are sparsely covered in time: for this object, there are a total of 45 observations over 4 years, which spans about 3000 typical RR Lyrae periods.
\label{PS1_multi-band_timeseries}}
\end{figure}

Even though the PS1 $3\pi$ survey holds a great potential for Galactic studies due to its depth and sky coverage, it is a challenging data set for selection of RR Lyrae stars due to its sparse temporal coverage, cadence, and asynchronous multi-band observations (see \autoref{PS1_multi-band_timeseries}). As we described in our previous work \citep{her16}, we overcame these challenges by characterizing time-series of PS1 sources using three statistics: a $\chi^2$-based variability indicator, a variability amplitude (in the $r_{\rm P1}$-band) $\omega_r$, and a variability time-scale $\tau$, where the latter two were obtained by fitting a damped random walk model to observed PS1 {\it multi-band} structure functions (see Section 3.2 of \citealt{her16}). When applied to the second internal PS1 data release (PV2), our approach yielded a candidate sample of $\sim150,000$ RR Lyrae stars covering three quarters of the sky and reaching up to 120 kpc from the Sun.

Building on the work by \citet{her16}, in this paper we use the final PS1 data release (PV3) to significantly increase the completeness and purity of the PS1 sample of RR Lyrae stars. Compared to \citet{her16}, we achieve higher completeness and purity by 1) having more observation epochs per object (72 in PV3 vs.~55 in PV2), 2) by excising fewer and thereby retaining more of these observations (using a machine-learning algorithm that more efficiently identifies bad photometric data), 3) by building a more detailed machine-learned model of RR Lyrae stars in data space, and 4) by developing and running CPU-intensive multi-band light curve fitting on PS1 time-series, thereby directly determining the RR Lyrae periods. The purer samples of RR Lyrae stars that this work delivers are especially important for studies of the Galactic halo (e.g., when searching for low-luminosity dwarf satellites), as stars incorrectly identified as RR Lyrae stars may cause appearance of spurious halo substructures \citep{ses10}.

\section{Data: PS1 $3\pi$ Light Curves }\label{Sec:PS1_3pi}

From an observational point of view, RR Lyrae stars are A-F type stars with distinct, periodically variable light curves. In the following sections, we describe data that capture these properties of RR Lyrae stars.

Pan-STARRS1 \citep[PS1]{kai10} is a wide-field optical/near-IR survey telescope system located at Haleakal\={a} Observatory on the island of Maui in Hawai`i. The largest survey undertaken by the telescope, the PS1 $3\pi$ survey \citep{cha11}, has observed the entire sky north of declination $-30\arcdeg$ in five filter bands \citep{stu10,ton12}, reaching $5\sigma$ single epoch depths of about 22.0, 22.0, 21.9, 21.0 and 19.8 magnitudes in $g_{\rm P1}$, $r_{\rm P1}$, $i_{\rm P1}$, $z_{\rm P1}$, and $y_{\rm P1}$ bands, respectively. The uncertainty in photometric calibration of the survey is $\la0.01$ mag \citep{sch12}, and the astrometric precision of single-epoch detections is 10 milliarcsec \citep{mag08}.

The PS1 $3\pi$ survey aimed to observe each position in two pairs of exposures per filter per year, where the observations within each so-called transit-time-interval pair were taken $\sim25$ minutes apart and in the same band. Thus, the survey should have obtained about 16 observations in each band (for a total of 80), but due to bad weather and telescope downtime, fewer epochs were observed ($\sim 70$ on average).  

Unlike \citet[see their Section 2.2]{her16}, we do not use bit-flags or other {\em ad hoc} procedures to exclude detections that may appear as non-astrophysical photometric outliers in PS1 time-series data (e.g., badly calibrated data, blended objects, etc.). We define a non-astrophysical photometric outlier as a photometric measurement that deviates by more than $2.5\sigma$ from its ``expected'' value, where $\sigma$ is the total photometric uncertainty of that detection. Instead, to identify and remove non-astrophysical photometric outliers, we employ a machine-learned model that uses other properties associated with a detection (e.g., its position on the chip, level of agreement with a Point-Spread-Function model, seeing, etc.) to {\em predict} whether a detection will be a $2.5\sigma$ outlier or not (Sesar et al., {\it in prep.}).

Validation tests have shown that our machine-learned outlier model identifies 80\% of all true $2.5\sigma$ outliers, and only misclassifies one good observation for every true $2.5\sigma$ outlier. For comparison, the outlier rejection approach adopted by \citet{her16} identifies almost all of the $2.5\sigma$ outliers, but it misclassifies eight good observations for every true $2.5\sigma$ outlier.

After removing photometric outliers from PV3 time-series (using our machine-learned outlier model), the average number of observations per object is 67 (out of the initial 72 observations). If we would have used the outlier rejection method of \citet{her16}, the number of observations per object would have decreased to $\sim30$.

To select objects with enough epochs for multi-band light curve fitting (\autoref{multi-band_light_curve_fitting}), and signal-to-noise ratios appropriate for variability studies, we require that PS1 light curves have (after outlier rejection):
\begin{itemize}
\item at least two epochs in $g_{\rm P1}$, $r_{\rm P1}$, and $i_{\rm P1}$ bands, and at least a total of two epochs in the ``red'' bands" ($z_{\rm P1}$ and $y_{\rm P1}$).
\item a total of at least 23 epochs,
\item and an uncertainty-weighted mean magnitude of $15 < \langle m \rangle < 21.5$ in at least one of the PS1 $g_{\rm P1}, r_{\rm P1}, i_{\rm P1}$ bands.
\end{itemize}

Dereddened optical colors are useful as they provide a rough estimate of the spectral type, and could help with the identification of RR Lyrae stars (which are A-F type stars). Thus, we correct observed PS1 magnitudes for extinction using the extinction coefficients of \citealt{sf11} (see their Table 6) and the \citealt{sch14} dust map and calculate $\langle g \rangle - \langle r\rangle$, $\langle r \rangle - \langle i\rangle$, $\langle i \rangle - \langle z\rangle$, $\langle z \rangle - \langle y\rangle$, and $\langle g \rangle - \langle i\rangle$ colors, where $\langle \rangle$ indicates an uncertainty-weighted mean magnitude. If for some reason an object is not observed in a particular PS1 band, the value of the color involving that band is reset to 9999.99.

To extract variability information from multi-band PS1 light curves, we calculate the variability indicator $\hat{\chi}^2$ (Equation 1 of \citealt{her16}), and fit a damped random walk model to PS1 multi-band structure functions. From the best-fit damped random walk model we measure the variability amplitude $\omega_r$ (in the $r_{\rm P1}$-band), and the variability time-scale $\tau$ (see Sections 3.1 to 3.3 of \citealt{her16}). As \citet{her16} have shown, these three parameters are very useful for separating different types of variable sources (e.g., quasars and RR Lyrae stars).

\section{Light Curve and Period Fitting}\label{Sec:lightcurve_fitting}

In this Section we describe several approaches to fitting the multi-band light curves, which will result in a period determination and a fit to the phased light curve.

\subsection{Multi-band Periodogram}\label{multi-band_periodogram}

A more detailed separation of variables can be obtained by studying the properties of phased (i.e., period-folded) light curves, such as the amplitude and shape \citep{ric11,dub11,elo16}. However, light curve folding requires an assumed or known period.

To measure the period of variability of a PS1 light curve, we use the multi-band periodogram of \citet{vi15} as implemented in \texttt{gatspy}, an open-source Python package for general astronomical time-series analysis\footnote{\url{http://www.astroML.org/gatspy/}} \citep{van15}. Briefly, the algorithm of \citet{vi15} models the phased light curves in each band as an arbitrary truncated Fourier series, with the period and optionally the phase, shared across all bands.

Since the phase offsets between RR Lyrae $griz$ light curves are smaller than 1\% \citep{ses10}, we adopt the \citet{vi15} shared-phase model when calculating the multi-band periodogram (i.e., we set \texttt{gatspy} parameters $N_{\rm base} = 1$ and $N_{\rm band} = 0$; see Section 5 of \citealt{vi15} for details). Furthermore, since RR Lyrae stars have periods between 0.2 and 0.9 days, we limit the period search to the same range. 

To test the accuracy of the \citet{vi15} period-finding algorithm on PS1 data, we use \texttt{gatspy} to calculate multi-band periodograms for 440 RR Lyrae stars previously studied by \citet{ses10}, but with the PS1 data at hand. From each periodogram, we select the period associated with the highest periodogram peak. We considered the periods measured by \citet{ses10} from more densely sampled Sloan Digital Sky Survey (SDSS; \citealt{yor00}) Stripe 82 observations to be ``true'' periods (see Section 2 of \citealt{ses10} for details on SDSS Stripe 82). We define a period to be accurately measured if the selected multi-band periodogram peak is within 2 sec of the true period.

We find that the \citet{vi15} period-finding algorithm accurately identifies the true period for 53\% of RR Lyrae stars observed by PS1 (37\% if the accuracy of 1 sec is required). Changing the \texttt{gatspy} $N_{\rm base}$ and $N_{\rm band}$ parameters did not significantly improve this result.

Using a mathematical, not physical, multi-band light curve model is one of the reasons why the \citet{vi15} period-finding algorithm fails to identify the true period for half of the RR Lyrae stars observed by PS1. When the light curves are sparse, there is no guarantee that the resulting best-fit multi-band model will be physical. If the light curve model is not constrained by external information on the physics of the problem at hand, inaccurate, or not robust, period estimates may result.

\begin{figure}
\plotone{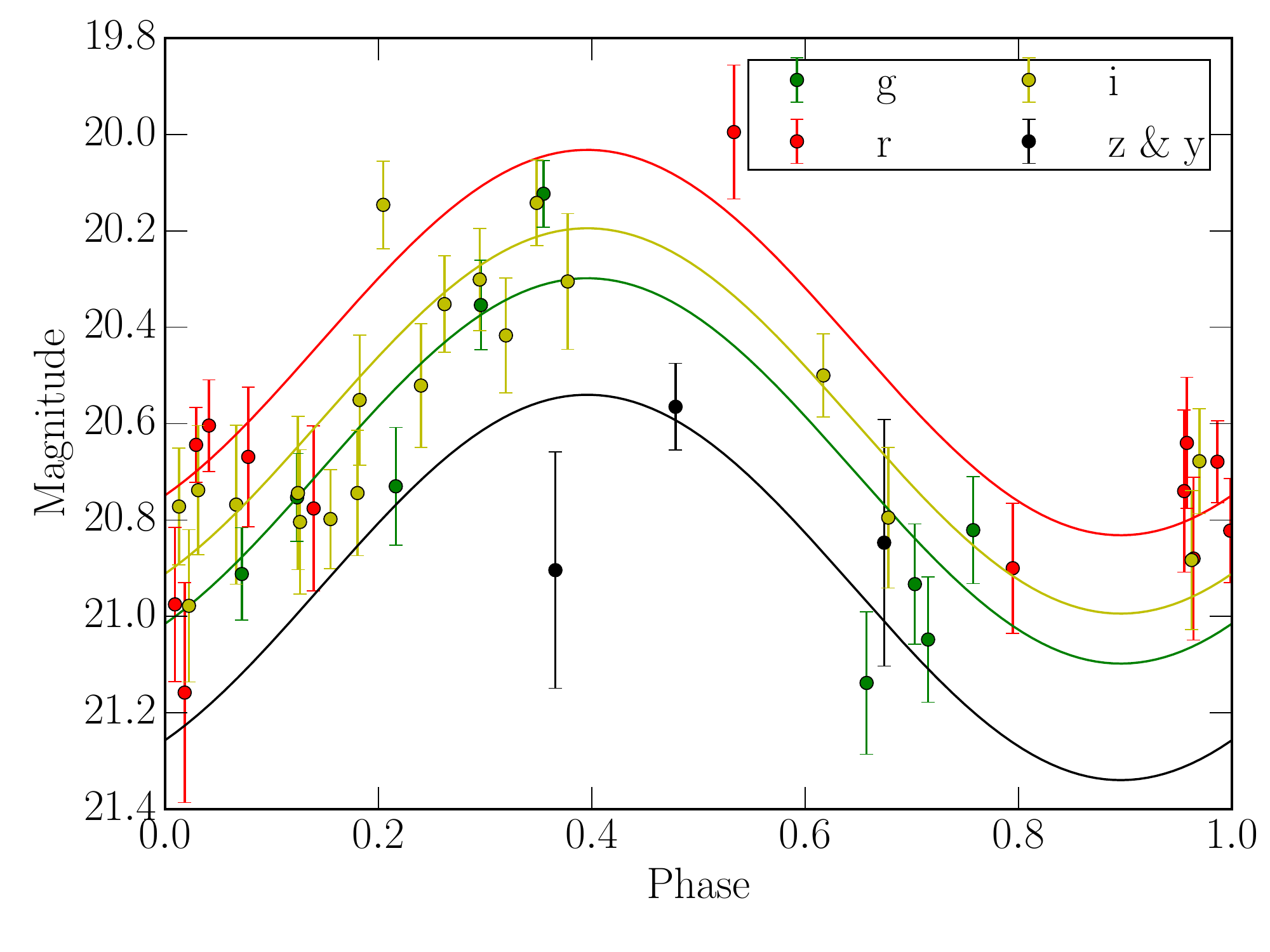}
\caption{
An unsuccessful attempt to accurately measure the period of the RR Lyrae star from \autoref{PS1_multi-band_timeseries}, using the multi-band periodogram of \citet{vi15}. Even though the best-fit multi-band model (sine curves) agrees with the phased PS1 light curves (symbols with errorbars) in the minimum $\chi^2$ sense, the period inferred by this modeling approach is incorrect. This happens because the algorithm permits light curve models that are not physical: the model's colors do not change as a function of phase.
\label{multi-band_fit}}
\end{figure}

An example of a best-fit, but non-physical (mathematical) multi-band model is shown in \autoref{multi-band_fit}. The $g-r$ color predicted by the best-fit model does not change as a function of phase (i.e., the difference between the green and red line), while in reality, it is well known that RR Lyrae stars have bluer $g-r$ color when they are brightest (e.g., \citealt{ses10}). Furthermore, the model predicts $i-z\sim0.2$ mag, while in reality $i-z\sim0$ mag. Due to a combination of a non-physical multi-band model and sparse PS1 data, the \citet{vi15} algorithm is unable to accurately measure the pulsation period of that particular star.

Since the \citet{vi15} algorithm fails to accurately measure the period for almost half of the RR Lyrae stars with the PS1 data at hand, we cannot use it to phase the light curves. However, we still calculate and use the multi-band periodogram in this work in the subsequent classification, as it improves the selection of RR Lyrae stars (see \autoref{second_classifier} below).

\subsection{Multi-band Light Curve Fitting and Periods}\label{multi-band_light_curve_fitting}

We now show that it is possible to accurately measure periods of RR Lyrae stars observed by PS1, by using a more realistic and physically constrained multi-band light curve model.

In principle, such a model could be obtained by extracting theoretical $griz$ light curves from pulsation models, such as those created by \citet{mar06}. However, a comparison of theoretical \citep{mar06} and empirical (i.e., observed)  SDSS $ugriz$ light curves by \citet[see their Figure 8]{ses10} has shown differences between the two light curve sets that cannot be explained by observational uncertainties. Due to these differences, we are reluctant to use theoretical multi-band models of RR Lyrae stars when measuring periods.

\begin{figure}
\plotone{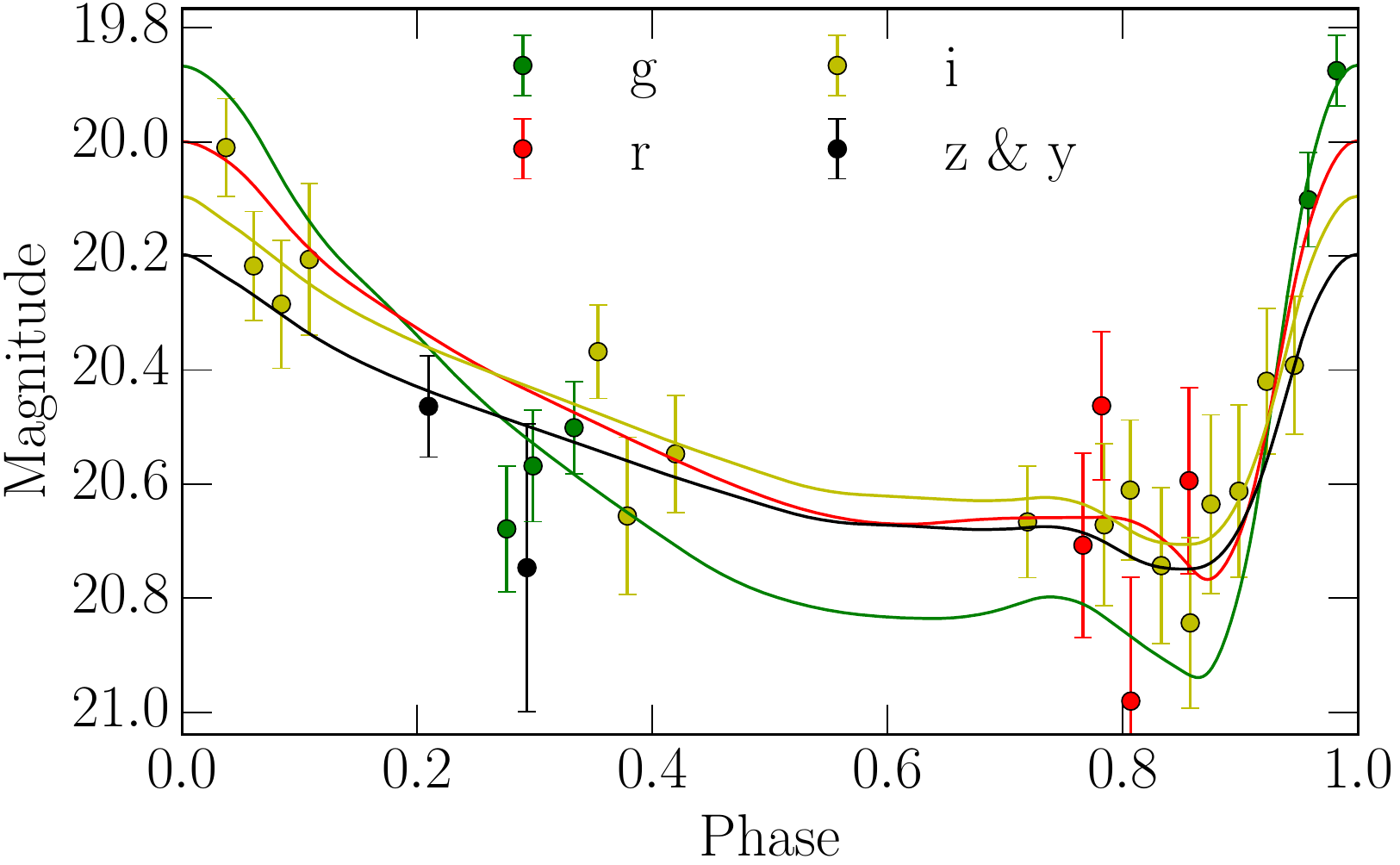}
\caption{
Phased PS1 light curves of the object shown in \autoref{PS1_multi-band_timeseries}, folded using the best-fit period measured from multi-band template fitting of PS1 data (see \autoref{multi-band_light_curve_fitting} for details). The best-fit multi-band template ($g,r,i,z\&y$) is overplotted. Even though this star is very faint ($r\sim20.5$ mag) and its light curve is sparsely sampled in PS1 (a total of 45 observations across 3000 periods), the period of $\sim0.51$ days measured using multi-band template fitting agrees within 2 sec with the value measured by \citet{ses10} from more densely sampled SDSS Stripe 82 data.
\label{template_fit}}
\end{figure}

Instead of using theoretical multi-band models, we adopt a set of 483 empirical $griz$ models. These models consist of $griz$ light curve templates that were fitted by \citet{ses10} to observed $griz$ light curves of 483 RR Lyrae stars in SDSS Stripe 82. The curves in \autoref{template_fit} illustrate one of the 483 empirical multi-band models. The set contains 379 type $ab$ multi-band templates, corresponding to RR Lyrae stars pulsating in the fundamental model (RRab stars), and 104 type $c$ multi-band templates, corresponding to RR Lyrae stars pulsating in the first overtone (RRc stars).

We define the $k$-th empirical multi-band model, which is a function of the pulsation phase $\phi$ as:
\begin{align}
g(\phi) &= FA_{\rm g}T_{\rm g}(\phi) + g_{\rm 0} - r_{\rm 0} + r^\prime\label{multi-band_templates_equations} \\ 
r(\phi) &= FA_{\rm r}T_{\rm r}(\phi) + r^\prime \nonumber \\
i(\phi) &= FA_{\rm i}T_{\rm i}(\phi) + i_{\rm 0} - r_{\rm 0} + r^\prime \nonumber \\
z(\phi) &= FA_{\rm z}T_{\rm z}(\phi) + z_{\rm 0} - r_{\rm 0} + r^\prime, \nonumber
\end{align}
where $T_{\rm m}(\phi)$ is the best-fit template light curve, $A_{\rm m}$ is the (known and fixed) amplitude of this template, and $m_{\rm 0}$ is the (known and fixed) best-fit magnitude at peak brightness (i.e., at $\phi = 0$) in the $m=g,r,i,z$ band of the $k$-th RR Lyrae star in SDSS Stripe 82 (see Table 2 of \citealt{ses10} for values of $A_{\rm m}$ and $m_{\rm 0}$). Note that the free parameter $r^\prime$ acts as a zero-point offset in our model, since the $griz$ light curves have been normalized by subtracting $r_{\rm 0}$ from each light curve. The free parameter $F$ allows the amplitudes of model $griz$ light curves to vary by up to 20\% from their original values (which are listed in Table 2 of \citealt{ses10}).

Qualitative inspection of phased PS1 $z$ and $y$ band light curves has shown that the two are roughly similar within photometric uncertainties. Therefore, we (can) treat all $y$-band observations as $z$-band observations in the remainder of the analysis.

Assuming a period of $P$ days and a phase offset $\phi_0$, we calculate the phase of each PS1 observation epoch as
\begin{equation}
\phi(t~|~ P, \phi_0) = \frac{(t - 2400000)\, {\rm modulo}\, P}{P} + \phi_0\label{phase},
\end{equation}
  where the time of observation $t$ is in units of heliocentric Julian days, and $-0.5 \leq \phi_0 < 0.5$. The purpose of the phase offset $\phi_0$ is to make sure the maximum light of the best-fit multi-band template occurs at $\phi = 0$. Note that the phase of an observation needs to be in the $0 \leq \phi < 1$ range. If it is outside of that range, one should add or subtract 1.

To find the best-fit values of $F$, $r^\prime$, $\phi_0$, and $P$ parameters for a given multi-band template, $k$, we minimize a $\chi^2$-like statistic calculated as
\begin{equation}
\chi^2_k = \sum_{m=g,r,i,z}\sum_{n=1}^{N_{\rm m,obs}}\left(\frac{m_{\rm m,n} - m_k\bigl (\phi(t_{\rm m,n}~|~ P, \phi_0)~|~F,r^\prime\bigr )}{\sigma_{\rm m,n}}\right)^2, 
\end{equation}
where $t_{\rm m,n}$, $m_{\rm m,n}$, and $\sigma_{\rm m,n}$ are the time, magnitude, and the photometric uncertainty of the $n$-th observation in the $m=g,r,i,z$ band (e.g., $t_{\rm g, 1}$, $g_{\rm 1}$, $\sigma_{\rm g, 1}$). The best-fit parameters (period, phase offset, etc.) are measured using the Differential Evolution algorithm of \citet{sp97} as implemented in \texttt{scipy}, an open-source Python package for scientific computing\footnote{\url{http://www.scipy.org}} \citep{scipy1, scipy2}.

\begin{figure}
\plotone{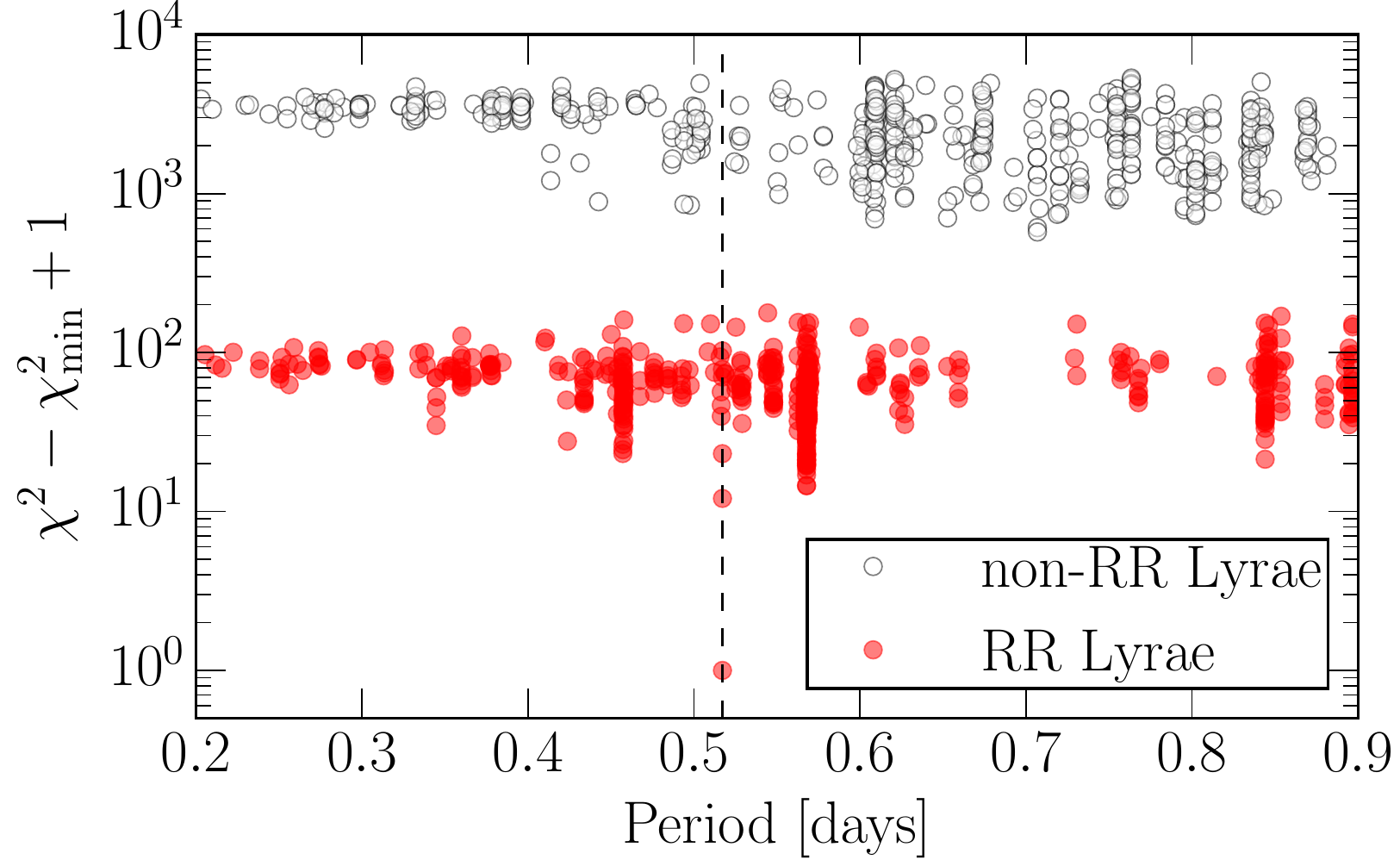}
\caption{
The symbols show best-fit periods and associated $\chi^2$ values obtained by fitting each of the 483 multi-band templates to PS1 light curves of a faint non-RR Lyrae object (open grey points), and the same faint RR Lyrae star shown in Figures~\ref{PS1_multi-band_timeseries} to~\ref{template_fit} (red points). The vertical dashed line shows the RR Lyrae star's true period measured by \citet{ses10} from SDSS Stripe 82 data. For this RR Lyrae star, the period associated with the best-fit template (i.e., the template with the smallest $\chi^2$ value) is consistent within 2 sec with the star's true period, indicating a successful period recovery. Note the classification power of the template fitting $\chi^2$ statistic, as it clearly separates the RR Lyrae star from the non-RR Lyrae object, even though both objects have similarly sampled PS1 light curves and signal-to-noise ratios.
\label{period_vs_chi2}}
\end{figure}

We perform multi-band light curve fitting in two runs. In the first run, we fit every multi-band template to a PS1 multi-band time-series, and for each template record the best-fit $\chi^2$ and model parameter values. When fitting a type $ab$ multi-band template, we constrain the minimization to periods ranging from 0.4 to 0.9 days. The minimization is constrained to periods ranging from 0.2 to 0.5 days when fitting a type $c$ multi-band template. The above period ranges are typical of type $ab$ (RRab) and type $c$ (RRc) RR Lyrae stars pulsating in the fundamental or the first-overtone mode, respectively. For illustration, \autoref{period_vs_chi2} shows the 483 best-fit $\chi^2_k$ and period values, one for each template light curve, measured for a faint RR Lyrae star and a faint non-RR Lyrae object. Clearly, there is a global minimum for the RR Lyrae light curve among the $\chi^2_k$ values.

In a second round of light curve fitting, we fit only type $ab$ or type $c$ templates, depending on the type of the best-fit template (i.e., the template with the lowest $\chi^2$ value) found during the first run. This time, the period range is restricted to $\pm2$ minutes around the period associated with the best-fit template from the first run. At the end of the second fitting iteration, we save only the best-fit $\chi^2$ and model parameter values associated with the best-fit template (of the second run).

\begin{figure}
\plotone{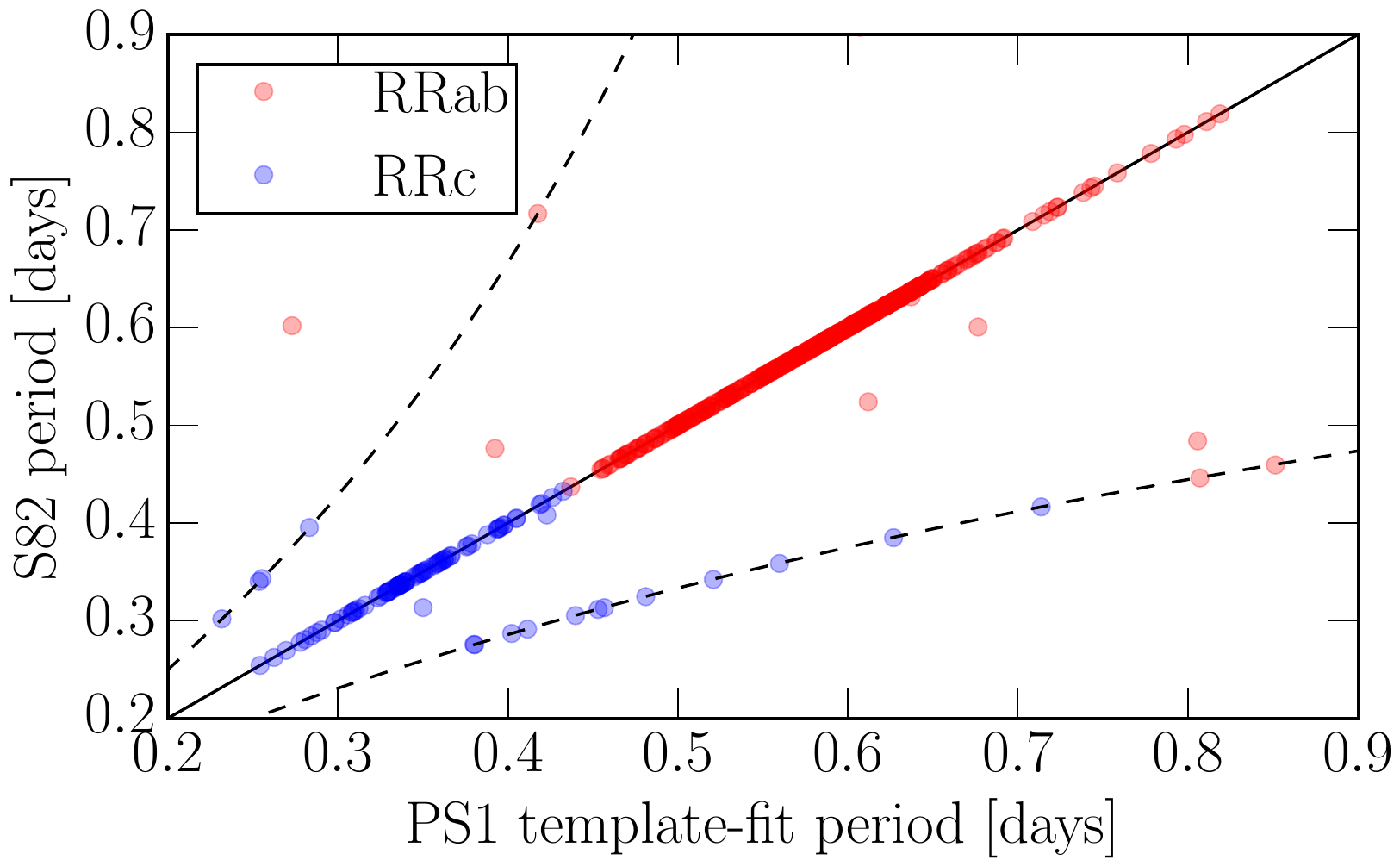}

\plotone{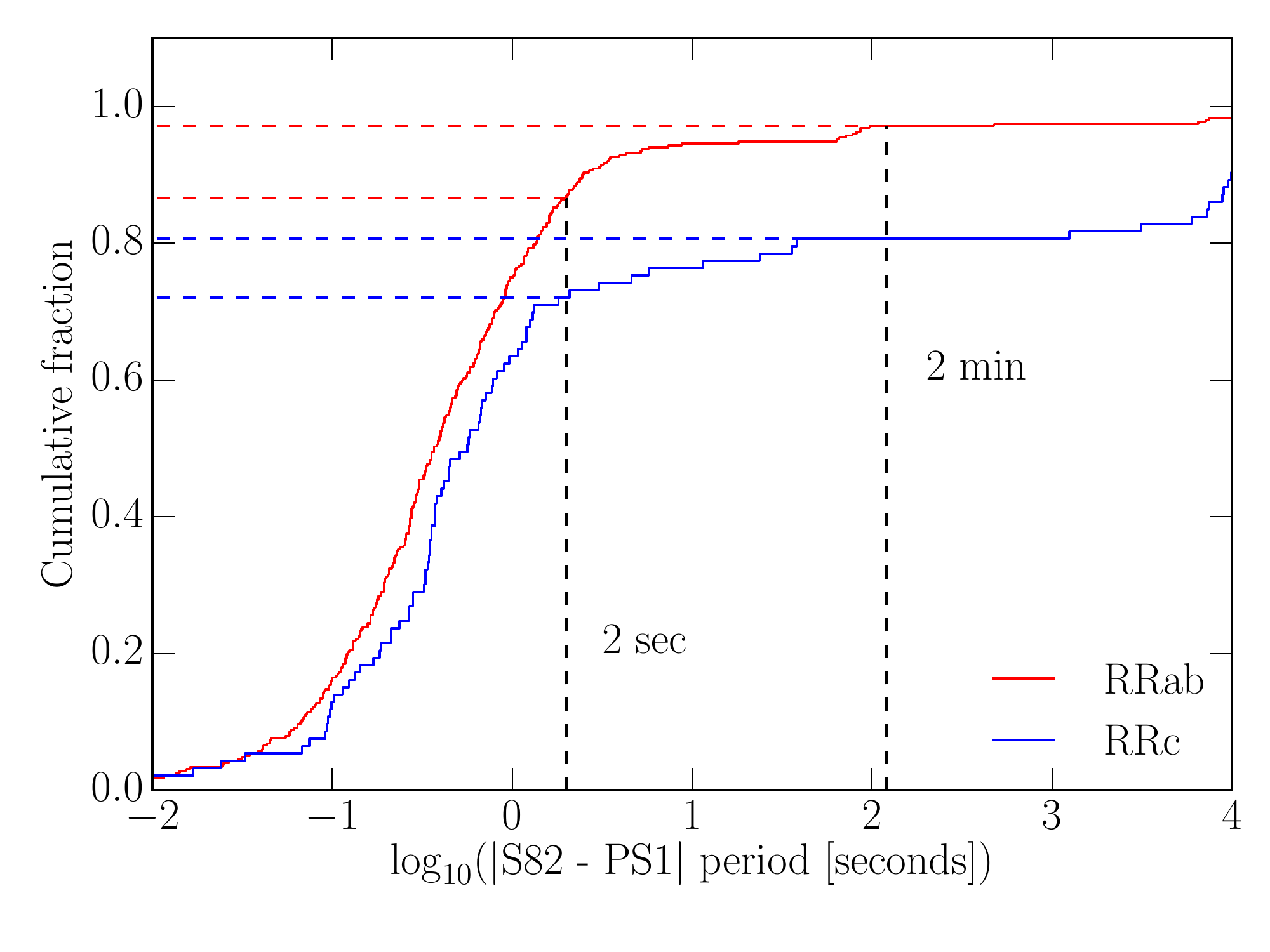}
\caption{
Accuracy, precision and robustness of the RR Lyrae period estimates obtained using our multi-band light curve template fitting of PS1 light curves. The top panel compares periods measured from Stripe 82 data (by \citealt{ses10}) with those measured from PS1 data using multi-band template fitting. The dashed lines show the 1-day beat frequency aliases. The bottom panel quantifies the period recovery: the period is accurately recovered (i.e., within 2 sec) for 87\% of RRab and 74\% of RRc stars. \label{period_comparison}}
\end{figure}

The result of applying this procedure to the PS1 lightcurves of 440 RR Lyrae stars in SDSS Stripe 82 is illustrated in \autoref{period_comparison}. Our multi-band template fitting method accurately measures periods for 85\% of RR Lyrae stars (87\% of RRab and 74\% of RRc stars), a 32\% improvement in period recovery over the \citet{vi15} algorithm. Within 1 sec, the period is recovered for 73\% of RR Lyrae stars (a 36\% improvement versus the \citealt{vi15} algorithm). If the period fitting returns a discrepant value, this can predominately be attributed to 1-day beat frequency aliasing (see \autoref{period_comparison}).

When fitting multiband templates to PS1 lightcurves of 440 RR Lyrae stars in SDSS Stripe 82, there is a possibility that some Stripe 82 RR Lyrae stars will be best fit with their own multiband templates (recall that multiband templates were constructed from observed SDSS light curves of individual Stripe 82 RR Lyrae stars). This ``self-fitting'' can be considered as a form of overfitting\footnote{Because the correct period is more likely to be found for a star when its own template is used.} which, if it happens frequently, may inflate our estimate of the accuracy of period recovery. However, only 6 out of 440 Stripe 82 RR Lyrae stars are fit with their own multiband templates, indicating that self-fitting does not happen frequently and that it does not significantly inflate our estimate of the accuracy of period recovery. The lack of self-fitting implies that many multi-band templates are quite similar to each other, and suggests that there is a potential for a computational speedup by removing redundant multi-band templates from the set.

As \autoref{period_vs_chi2} illustrates, the multi-band light curve fitting also provides useful information for separating RR Lyrae stars and non-RR Lyrae objects. For example, the average and best-fit $\chi^2$ values measured for the RR Lyrae star are vastly lower than the corresponding values measured for the non-RR Lyrae object, even though both objects have PS1 light curves of similar signal-to-noise ratio and sampling. This result is not unexpected, since $\chi^2$ measures the statistical agreement between the (observed) phased PS1 light curve and a best-fit empirical multi-band light curve model of an RR Lyrae star, and thus quantifies how ``RR Lyrae-like'' an object is.

In order to further characterize how RR Lyrae-like a PS1 multi-band light curve is, we measure additional properties of phased PS1 light curves, such as the entropy of the phased light curve, the \citet{ste96} $J$ index, as well as $\sim20$ other properties (called {\em features} in machine learning), and describe them in more detail in \autoref{extraction_of_features}.

\subsection{Resulting RR Lyrae Distance Precision}\label{distance_precision}

Along with measuring an accurate period ($87\%$ of the time for RRab stars), an important aspect of multi-band light curve fitting is also the increased precision in estimating the star's average flux (or magnitude). As RR Lyrae stars follow a tight period-absolute magnitude-metallicity (PLZ) relation, we show that PS1 data constrain distances of RR Lyrae stars with a 3\% precision, even if the metallicity of an RR Lyrae star is unknown.

Theoretical and empirical studies \citep[e.g.,][]{cat04,mar15,sol06,bra15} have shown that the absolute magnitudes of RR Lyrae stars can be modeled as
\begin{equation}
M = \alpha\log_{\rm 10}(P/P_{\rm ref}) + \beta({\rm [Fe/H]} - {\rm [Fe/H]_{\rm ref}}) + M_{\rm ref} + \epsilon,\label{PLZ}
\end{equation}
where $P$ is the period of pulsation, $M_{\rm ref}$ is the absolute magnitude at some reference period $P_{\rm ref}$ and metallicity ${\rm [Fe/H]_{\rm ref}}$ (here chosen to be $P_{\rm ref}=0.6$ days and ${\rm [Fe/H]_{\rm ref}}=-1.5$ dex), and $\alpha$ and $\beta$ describe the dependence of the absolute magnitude on period and metallicity, respectively. The $\epsilon$ is a standard normal random variable with mean 0 and standard deviation $\sigma_{\rm M}$, that models the intrinsic scatter in the absolute magnitude convolved with unaccounted measurement uncertainties.

To constrain the PLZ relations for RRab stars in PS1 bandpasses, we use a probabilistic approach described in detail in \autoref{PLZ_derivation}, where the data include metallicities and distance moduli of PS1 RR Lyrae stars in five Galactic globular clusters. The end product of this approach is a joint posterior distribution of all model parameters. To describe the marginal posterior distributions of individual model parameters, we measure the median, the difference between the 84th percentile and the median, and the difference between the median and the 16th percentile of each marginal posterior distribution (for a Gaussian distribution, these differences are equal to $\pm1$ standard deviation). We report these values in \autoref{PS1_PLZ}.

\capstartfalse
\begin{deluxetable*}{ccccc}
\tablecolumns{5}
\tablecaption{PLZ Relations for PS1 bandpasses\label{PS1_PLZ}}
\tablehead{
\colhead{Band} & \colhead{$\alpha$} & \colhead{$\beta$} & \colhead{$M_{\rm ref}$} & \colhead{$\sigma_{\rm M}$} \\
\colhead{ } & \colhead{(mag dex$^{-1}$)} & \colhead{(mag dex$^{-1}$)} & \colhead{(mag)} & \colhead{(mag)}
}
\startdata
$g_{\rm P1}$ & $-1.7\pm0.3$ & $0.08\pm0.03$ & $0.69\pm0.01(rnd)\pm0.03(sys)$ & $0.07\pm0.01$ \\
$r_{\rm P1}$ & $-1.6\pm0.1$ & $0.09\pm0.02$ & $0.51\pm0.01(rnd)\pm0.03(sys)$ & $0.06\pm0.01$ \\
$i_{\rm P1}$ & $-1.77\pm0.08$ & $0.08\pm0.02$ & $0.46\pm0.01(rnd)\pm0.03(sys)$ & $0.05\pm0.01$ \\
$z_{\rm P1}^a$ & $-2.2\pm0.2$ & $0.06\pm0.02$ & $0.46\pm0.01(rnd)\pm0.03(sys)$ & $0.05\pm0.01$
\enddata
\tablenotetext{a}{The PLZ relation for the $z_{\rm P1}$ band was derived using $z_{\rm P1}$ and $y_{\rm P1}$ band observations, since a qualitative inspection of phased PS1 $z$ and $y$ band light curves has shown that the two are roughly similar within photometric uncertainties.}
\end{deluxetable*}
\capstarttrue

Overall, the PLZ relations behave as expected: as the bandpass moves to redder wavelengths, the dependence on the period increases (i.e., the $\alpha$ parameter becomes more negative), the dependence on the metallicity (i.e., the $\beta$ parameter) and the scatter $\sigma_{\rm M}$ decrease, and the reference absolute magnitude becomes brighter. Similar trends were also observed in previous theoretical and observational studies \citep[e.g.,][]{mar15,bra15}. Since the PLZ relation for the $i_{\rm P1}$ band is most tightly constrained and has low metallicity dependence, we use it hereafter when measuring distances.

The metallicity information is not available for the vast majority of stars in PS1. To estimate the uncertainty in absolute $i_{\rm P1}$ magnitudes when the metallicity is unknown, we assume that RR Lyrae stars are drawn from the halo metallicity distribution function, represented with a Gaussian distribution centered on $-1.5$ dex and with a standard deviation of 0.3 dex \citep{tomoII}. The resulting uncertainty in $M_{\rm i_{\rm P1}}$ is then $\sigma_{\rm M_{\rm i_{\rm P1}}}=0.06(rnd) \pm0.03(sys)$ mag, and the expression for $M_{\rm i_{\rm P1}}$ simplifies to 
\begin{equation}
M_{\rm i_{\rm P1}} = -1.77\log_{\rm 10}(P/0.6) + 0.46.\label{abs_mag_i_band}
\end{equation}

To calculate distance moduli of PS1 RR Lyrae stars, we use flux-averaged $i_{\rm P1}$-band magnitude and \autoref{abs_mag_i_band}. For the uncertainty in distance modulus, we adopt $\sigma_{\rm DM} = \sigma_{\rm M_{\rm i_{\rm P1}}}=0.06(rnd) \pm0.03(sys)$ mag. This corresponds to a distance precision of $\sim3\%$, as long as dust extinction is not an important issue.

To validate \autoref{abs_mag_i_band}, we compute median distance moduli for three dwarf spheroidal galaxies, using the PS1 data for their RR Lyrae stars. We find $DM=19.51\pm0.03$ mag for Draco, $DM=19.67\pm0.03$ mag for Sextans, and $DM=19.17\pm0.03$ mag for Ursa Minor, where the uncertainty in distance moduli is dominated by the systematic uncertainty in absolute magnitude. The values for Draco and Sextans agree well with the literature values of $19.40\pm0.17$ and $19.67\pm0.1$ \citep{bon04,lee09}, respectively. The $DM=19.11\pm0.03$ mag we measure for Ursa Minor agrees with the $19.18\pm0.12$ mag measurement of \citet{mb99}, but disagrees with the $19.4\pm0.1$ mag value measured by \citet{car02}. The rms scatter of distance moduli of RR Lyrae stars in these dwarf galaxies is $\sigma_{DM}\approx 0.05$ mag. This scatter empirically verifies the intrinsic scatter of $\sigma_{\rm M}=0.05$ mag we measured when fitting the PLZ relation in the $i_{\rm P1}$ band.

\subsection{WISE Data}\label{wise}

Quasars (QSOs) are one of the biggest sources of contamination when selecting RR Lyrae stars, especially at faint magnitudes (i.e., as the probed volume of the Universe increases). They overlap with RR Lyrae stars in $g-r$ and redder optical colors (e.g., Figure 4 of \citealt{ses07}), and may look as variable as RR Lyrae stars when observed in sparse datasets (such as PS1, Figure 3 of \citealt{her16}). Because most QSOs have a hot dust torus, they show an excess of radiation in the mid-infrared part of the spectrum, and have the WISE \citep{wri10} mid-infrared color $W12=W1-W2 > 0.5$ mag (Figure 2 of \citealt{nik14}). RR Lyrae stars, on the other hand, have $W12 \sim 0$ mag.

To better separate QSOs and RR Lyrae stars, we supplement PS1 data with the $W12$ color provided by the all-sky WISE mission by matching PS1 and WISE positions using a $1\arcsec$ radius. If a PS1 object does not have a WISE $W1$ or $W2$ measurement, or those measurements have uncertainties greater than 0.3 mag (i.e., the WISE detection is less than $5\sigma$ above the background), we reset its $W12$ color to 9999.99 (this happens for about 50\% of objects). We also calculate the $\langle i\rangle - W1$ color, and set its value to 9999.99 if one of its magnitudes is missing.

\section{RR Lyrae Identification}\label{Sec:Method}

We wish to build a model that returns the probability of an object to be an RR Lyrae star, given the data from \autoref{Sec:PS1_3pi} and the light curve fits from \autoref{Sec:lightcurve_fitting}. 
We will address this problem with a supervised machine learning approach, where we use a training set (labeled or classified objects and their data) to infer a function that determines the class of unlabeled objects from their data. Since we have a reliable training set (see \autoref{training_set}), supervised learning techniques (\autoref{classifiers}) represent a natural choice for building a classification model for a selection of RR Lyrae stars. The light curve fitting and its $\chi^2$ value (\autoref{multi-band_light_curve_fitting}) play a crucial role in this process.

\subsection{Training Set}\label{training_set}

To ``learn'' how to classify objects, supervised algorithms need to be trained using a subset of the data in which each object is labeled, i.e. their class is known in advance.

Our main training set consists of 1.9 million PS1 objects located in the SDSS Stripe 82 region ($310\arcdeg > RA < 59\arcdeg$, $|Dec| < 1.25\arcdeg$) that are brighter than 21.5 mag, have at least 23 observations (\autoref{Sec:PS1_3pi}), and are at least $24\arcmin$ (2 tidal radii; \citealt{har96} (2010 edition)) away from the center of globular cluster NGC 7089. To label the objects in the training set, we match them to \citet{ses10} and \citet{suv12} catalogs of RR Lyrae stars. If the position of an RR Lyrae star in one of these two catalogs matches the position of the closest PS1 object within $1\arcsec$, we label the PS1 object as an ``RR Lyrae'' (class 1; there are 462 such matches, and only 3 RR Lyrae stars do not have a PS1 match). The remaining PS1 objects are labeled as ``non-RR Lyrae'' (class 0). Out of 465 matched RR Lyrae stars, we know that 364 are of type $ab$ (RRab) and 98 are of type $c$ (RRc).

Since the SDSS Stripe 82 observations are slightly deeper and have 6 times more epochs than PS1 observations, we consider the \citet{ses10} and \citet{suv12} RR Lyrae catalogs to be 100\% pure and complete up to the adopted faint PS1 magnitude limit (and likely beyond). Consequently, we consider the labels of PS1 objects in SDSS Stripe 82 as the ``ground truth'' when measuring the efficiency of our selection method (i.e., the selection completeness and purity).

In SDSS Stripe 82, the majority of previously identified RR Lyrae stars are located within $\sim30$ kpc of the Sun (400 stars or $83\%$ of the sample, see Figure 10 of \citealt{ses10}), and are thus fairly bright ($r_{\rm P1} < 18.5$). This distribution is the result of Galactic structure, and is not a selection effect. To enhance the training set with fainter, and thus more distant RR Lyrae stars, we use RR Lyrae stars in the Draco dwarf spheroidal galaxy, located at a heliocentric distance of $\sim80$ kpc \citep{kin08}. If the position of an RR Lyrae star in the \citet{kin08} catalog matches the position of the closest PS1 object (with at least 23 observations) within $1\arcsec$, we label the PS1 object as a ``RR Lyrae'' (there are 261 such matches, and 5 RR Lyrae stars do not have a PS1 match). Out of 261 matched RR Lyrae stars, 205 are of type $ab$ (RRab), 30 are of type $c$ (RRc), and 25 are $d$-type RR Lyrae stars (RRd) that pulsate simultaneously in the fundamental mode and first overtone.

\subsection{Supervised Learning}\label{classifiers}

To build this classification model we use \texttt{XGBoost}\footnote{\url{https://github.com/dmlc/xgboost}} \citep{xgb}, an open-source implementation of the {\em gradient tree boosting} supervised machine learning technique \citep{fri01}.

We use gradient tree boosting because the technique produces a prediction model in the form of an ensemble of decision trees, and because tree-based models are robust to uninformative features\footnote{In machine learning, a feature is an individual measurable property of a phenomenon being observed (e.g., period of variability, color, light curve amplitude, etc.).} \citep{has09,ric11,dub11}. This fact supports the usage of a large number of features when building the classification model, even when some of them may not be useful. By permitting the classification algorithm to consider even seemingly uninformative features, we allow it to consider potential correlations between data that may improve the classification in the end.

Given the resilience of gradient tree boosting to uninformative features, and the improvement in classification that additional features may bring, the best approach seems to be to train the classifier using the full set of features. However, this is impractical for the data set at hand. While calculating mean optical colors, low-level variability statistics (\autoref{Sec:PS1_3pi}), and the multi-band periodogram takes less than a second per object, multi-band light curve fitting takes $\sim30$ min per object. Given that our training set contains about 1.9 million objects, calculating all these features for all objects in the training set would be computationally prohibitive.

Instead of training a single classifier using the full set of features, we build three progressively more detailed classifiers using progressively smaller, but purer training sets (purer in the sense that the fraction of RR Lyrae stars in the training set increases). We describe these classifiers in Sections~\ref{first_classifier} to~\ref{third_classifier}, but first give an overview of how a classifier is trained in \autoref{classifier_training}

\subsection{Overview of Classifier Training}\label{classifier_training}

In brief, the steps in training a classifier are to:
\begin{enumerate}
\item Select training objects and input features.
\item Tune \texttt{XGBoost} hyperparameters and train the classifier.
\item Measure the classification performance with a purity vs.~completeness curve.
\end{enumerate}
While the first step is fairly self-explanatory, the remaining steps require further explanation.

The gradient tree boosting technique produces a prediction model in the form of an ensemble of decision trees. The number of trees in the ensemble, the maximum depth of a tree, the fraction of features that are considered when constructing a tree, and many other parameters that affect the manner in which the trees are grown and pruned, can be controlled via parameters\footnote{See \url{https://xgboost.readthedocs.io/en/latest//parameter.html} for the full list.} exposed by the \texttt{XGBoost} package. By properly tuning these model {\em hyperparameters}, we can ensure that the classification produced by the model is not sub-optimal (e.g., not overfitted).

Before tuning the hyperparameters, we select input features, and shuffle and split the training set into two equal-sized sets which we call {\em development} and {\em evaluation} sets. We use stratified splitting, i.e., we make sure that the ratio of RR Lyrae and non-RR Lyrae objects is equal in both sets.

To find the optimal hyperparameters, we use the \texttt{GridSearchCV} function in the \texttt{scikit-learn} open-source package for machine learning\footnote{\url{http://scikit-learn.org}} \citep{scikit-learn}. \texttt{GridSearchCV} selects test values of hyperparameters from a grid, and then measures the performance of the classification model (for the given hyperparameters) using ten-fold stratified cross-validation on the development set. In detail, the development set is split into ten subsets (using stratified splitting, see above), the model is trained on nine subsets, and the probability of being an RR Lyrae star\footnote{Computed as the mean predicted class probabilities of the trees in the forest. Given a single tree, the probability that an object is of the RR Lyrae class (according to that tree) is equal to the fraction of training samples of the RR Lyrae class in the leaf in which the object ends up.} (hereafter, the classification score\footnote{In this work, we use three classifiers and label their scores as $score_{\rm j}$, where $j=1,2,3$.}) is obtained from the trained model for objects in the tenth (i.e., withheld) subset. The performance of the classification is evaluated on the withheld set using some metric (see Sections~\ref{first_classifier} to~\ref{third_classifier} for details), and the whole procedure is repeated nine more times, each time with a different withheld set. The average of the ten performance evaluations is stored, and the set of hyperparameters with the best average performance is used when training the classifier (step 3).

To verify whether the choice of the development set significantly affects the tuning of hyperparameters, we evaluate the performance of the tuned model on the evaluation set (which was not used by \texttt{GridSearchCV} during hyperparameter optimization), and then repeat the tuning process, but this time we use the evaluation set for tuning and the development set for evaluation. We find that the tuning procedure returns similar values of hyperparameters, regardless of the choice of the development set, indicating that the tuning of hyperparameters is not significantly biased by our choice of the development set.

\begin{figure}
\plotone{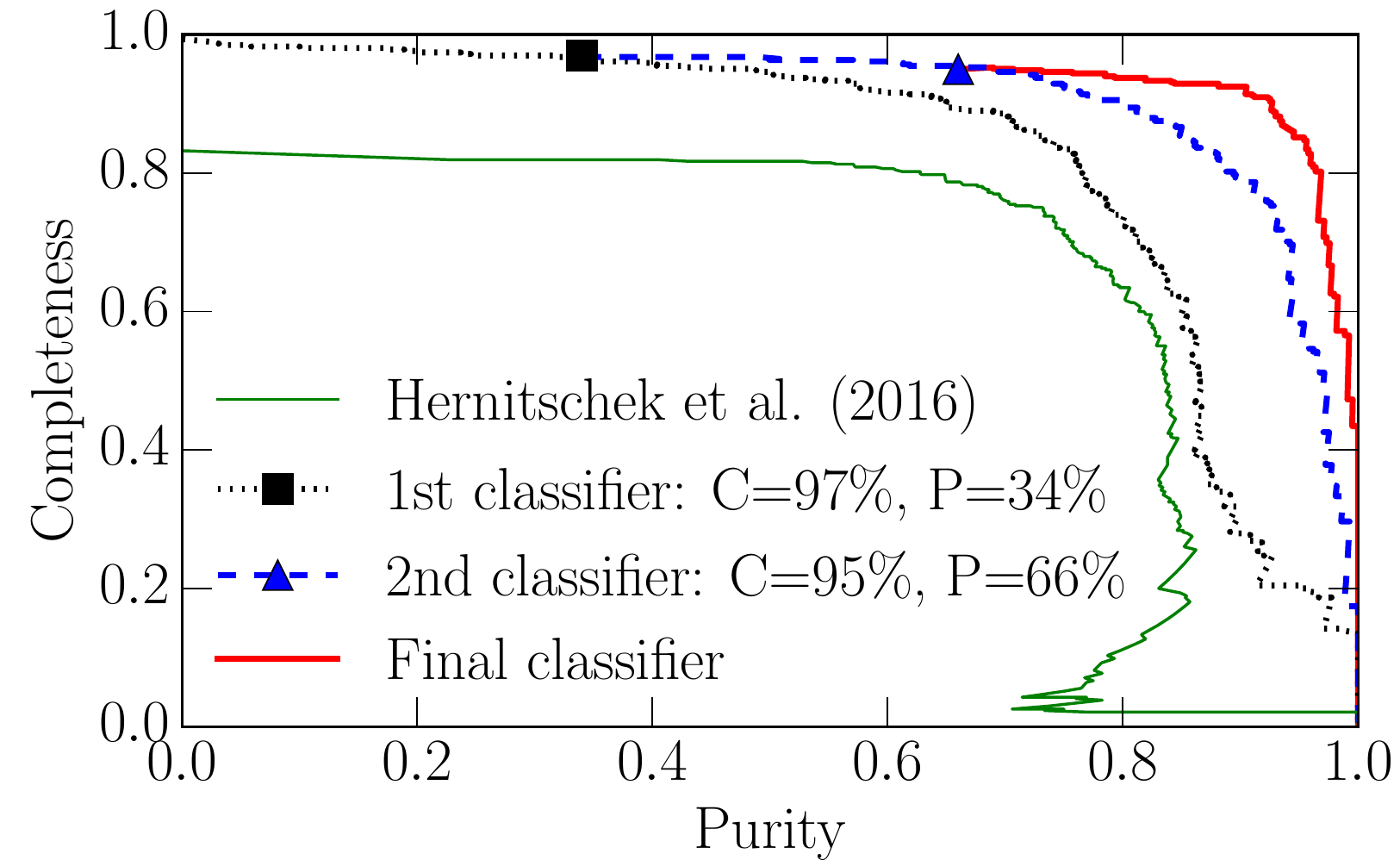}
\caption{
The power of the multi-band light curve fitting in the classification of RR Lyrae stars. The figure shows purity vs.~completeness curves produced by progressively more detailed classifiers described in Sections~\ref{first_classifier} to~\ref{third_classifier}. The ideal classifier should approach the top right corner of the diagram. The square and star symbols show the purity and completeness of the classification with the adopted choice of scores returned by the first and second classifier ($score_{\rm 1} >0.01$ and $score_{\rm 2} > 0.13$, respectively). The initial completeness is 99\% due to initial data quality cuts (\autoref{Sec:PS1_3pi}). Using the final classifier, we can select samples of RR Lyrae stars that are, for example, 90\% complete and 90\% pure.  
\label{purity_completeness_curves}}
\end{figure}

Once the hyperparameters are tuned and the classifier is trained, we evaluate the performance of the classification using a purity vs.~completeness curve (see \autoref{purity_completeness_curves} for examples). To measure this curve, we use the classification scores of objects in the Stripe 82 part of the training set (see \autoref{training_set} for details on this set). The classification scores of these objects were calculated using the ten-fold cross-validation on the full (Stripe 82 and Draco) training set. For any threshold on the score and knowing the true class of each object in SDSS Stripe 82, we obtain the fraction of recovered RR Lyrae stars (completeness), and the fraction of RR Lyrae stars in the selected sample (purity).

\subsection{First Classification Step: Optical/IR Colors and Variability}\label{first_classifier}

To train the first classifier, we use the full training set of 1.9 million objects (\autoref{training_set}), and adopt their variability statistics, as well as average PS1 and WISE colors, as input features for the classifier (for a total of 10 features, see Sections~\ref{Sec:PS1_3pi} and~\ref{wise} for details); we do not use the light curve fitting of Section~\ref{Sec:lightcurve_fitting}. When tuning the classifier, we select the values of hyperparameters that maximize the area under the purity vs.~completeness curve. The black dotted line in \autoref{purity_completeness_curves} characterizes the performance of the trained classifier.

Our first classification outperforms the one obtained by \citet{her16} for all choices of sample purity and completeness (i.e., for all thresholds on the classification score). This is attributable perhaps foremost to a substantially greater number of observations per object in our dataset (67 vs.~35 in \citealt{her16}). The hyperparameter tuning, and use of a different machine learning algorithm (\texttt{XGBoost} vs.~\texttt{scikit-learn} Random Forest) may also contribute to better performance.

 Using a cut on the classification score of $score_{\rm 1} > 0.01$ ($score_{\rm 1}$), we are able to reduce the number of objects under consideration by more than three orders of magnitude (from about 1.9 million to $\sim1500$), while losing only $2\%$ of RR Lyrae stars. However, the purity of the selected sample is still unacceptably low (only 34\%). In order to improve the purity of the selected sample we need to train the classifier using additional features, such as the multi-band periodogram (\autoref{multi-band_periodogram}).

\subsection{Second Classification Step: Multi-band Periodogram}\label{second_classifier}

For 1568 objects that pass the first classification cut ($score_{\rm 1} > 0.01$) we calculate multi-band periodograms and extract top 20 periods from each periodogram. Along with the periods, we also extract the power of each period (i.e., the height of the periodogram at that period). 

\begin{figure}
\plotone{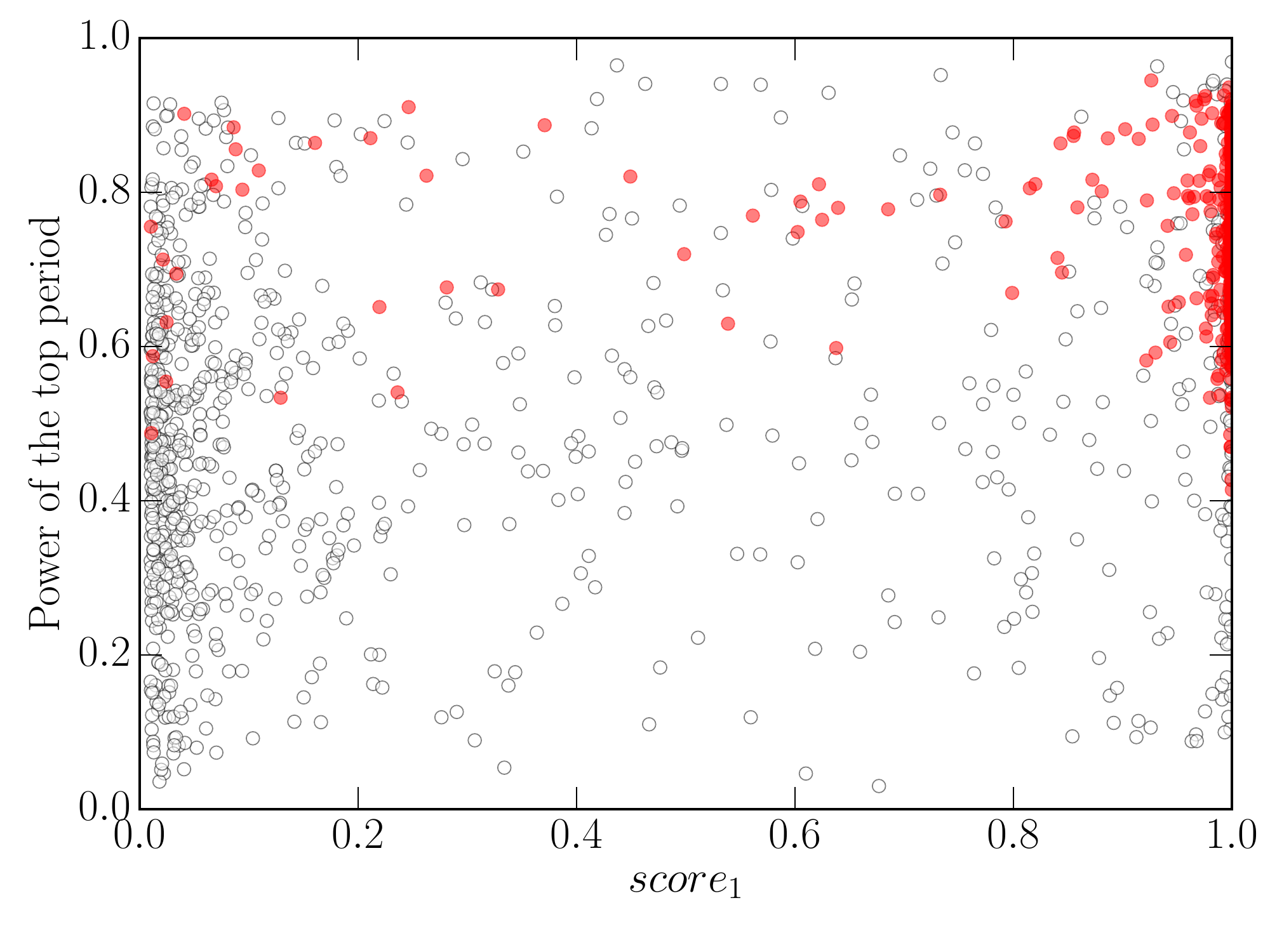}
\caption{
This plot shows that the multi-band periodogram contains useful information for separating RR Lyrae stars (solid circles) from non-RR Lyrae objects (open circles). Even though some true RR Lyrae stars may have low $score_{\rm 1}$ values (i.e., are not recognized by the first classifier as likely RR Lyrae stars), the power of the top period clearly separates them from non-RR Lyrae objects (e.g., $power_{\rm 0} \ga0.4$).
\label{power_vs_score}}
\end{figure}

As \autoref{power_vs_score} illustrates, the multi-band periodogram contains useful information for separating RR Lyrae stars from non-RR Lyrae objects. In principle, we could improve the purity of the selection by simply keeping all objects with $power_{\rm 0} > 0.4$, without a loss of completeness. On the other hand, we may achieve even better classification if we provide the {\em entire} set of periods and their powers to \texttt{XGBoost}, and let the algorithm decide which features to use.

To improve the classification of RR Lyrae stars, we create a new feature set by combining 10 features used by the first classifier, with the top 20 periods and their powers obtained from the multi-band periodogram (for a total of 50 features). As the training set, we use $\sim1500$ objects remaining from the initial training set after the $score_{\rm 1} > 0.01$ cut. When tuning hyperparameters, we adopt values that optimize the area under the purity vs.~completeness curve. The blue dashed line in \autoref{purity_completeness_curves} characterizes the performance of the second classifier.

We find that the addition of multi-band periodogram data improves the selection of RR Lyrae stars, as evidenced by higher sample purity at a given completeness (blue dashed line in \autoref{purity_completeness_curves}). For example, at 90\% completeness, adding multi-band periodogram data increases the purity of the selected sample by 15\% (with respect to the purity delivered by the first classifier, black dotted line).

\subsection{Final Classification Step: Multi-band Light Curve Fitting}\label{third_classifier}

Given the information in hand, a nearly optimal classification may be obtained by also including features extracted from phased multi-band light curves (\autoref{multi-band_light_curve_fitting} and~\autoref{extraction_of_features}), but first, we need to fit multi-band light curves to objects under consideration.

Since multi-band light curve fitting is computationally quite expensive ($\sim30$ min per CPU and per object), we only do it for 910 training objects that have $score_{\rm 2} > 0.13$, where $score_{\rm 2}$ is the classification score produced by the second classifier. According to \autoref{purity_completeness_curves}, this selection cut returns a sample with 66\% purity and 95\% completeness. By using this cut, we avoid fitting objects that are not likely to be RR Lyrae stars (i.e., we do not waste CPU time), but at the same time, do not reject many true RR Lyrae stars (i.e., the completeness decreases by 2\% due to this cut, with respect to the 97\% completeness obtained after the $score_{\rm 1} > 0.01$ cut). In principle, we could have reached the same sample purity using a cut on the first classification score ($score_{\rm 1}$), but the decrease in completeness would have been much greater (a 6\% decrease).

In Sections~\ref{first_classifier} and~\ref{second_classifier}, we have trained classifiers using non-RR Lyrae objects (class 0) and RR Lyrae stars (class 1), that is, we have performed {\em binary} classifications. The reason for this two-step procedure was practical. In order to make the multi-band template fitting computationally feasible, we had to reduce the number of objects under consideration to a manageable level (by increasing the purity of the selected sample), while retaining as many true RR Lyrae stars as possible (i.e., by keeping the selection completeness as high as possible). Doing this required only knowing whether an object is likely an RR Lyrae star or not. By using cuts on $score_{\rm 1}$ and $score_{\rm 2}$ (binary) classification scores, we were able to reduce the number of objects from about 1.9 million to 900, while retaining $95\%$ of RR Lyrae stars.

We now take a further step, determining whether an object is a non-RR Lyrae object (class 0), a type $ab$ (class 1), or a type $c$ or $d$ RR Lyrae star (class 2) through a {\em multiclass} classification.

To train the final (multiclass) classifier, we use 910 training objects that have $score_{\rm 2} > 0.13$ (and of course, $score_{\rm 1} > 0.01$). Of these 910 objects, 541 are RRab stars, and 144 are RRc or RRd stars (based on \citealt{ses10} and \citealt{kin08} classifications). The remaining 225 objects are non-RR Lyrae objects. The feature set consists of 50 features employed by the second classifier, and 20 features extracted from phased multi-band light curves (\autoref{extraction_of_features}). Since we are training a multiclass classifier, when tuning hyperparameters we adopt values that minimize the logistic (or cross-entropy) loss \citep{PRML}. The thick red line in \autoref{purity_completeness_curves} characterizes the purity and completeness of the selection as a function of the threshold on $score_{\rm 3,ab} + score_{\rm 3,c}$, where $score_{\rm 3,ab}$ and $score_{\rm 3,c}$ are RRab and RRc classification scores, respectively.

\section{Verification and Analysis of the RR Lyrae Selection at High Galactic Latitudes}

The purity vs.~completeness curve obtained using the full classifier (the thick red line in \autoref{purity_completeness_curves}) shows that we can select samples of RR Lyrae stars that are 90\% complete and 90\% pure; we deem this a gratifying and impressive success. However, it is important to keep in mind that the purity and completeness shown in \autoref{purity_completeness_curves} are {\em integrated} over RRab and RRc stars, and over a range of distances (or magnitudes; roughly, 5 to 120 kpc, $14.5 < \langle r \rangle/{\rm  mag} < 21$). Since we have reasons to expect variations in purity and completeness as a function of type and distance (e.g., because the classification becomes more uncertain as objects get fainter and light curves become noisier), and because the knowledge of such variations is important for studies of Galactic structure (e.g., when measuring the stellar number density profile), below we present a more detailed analysis of the selection of RR Lyrae stars at high galactic latitudes ($|b| > 15\arcdeg$).

\subsection{Purity and Completeness in Detail}\label{section_purity_completeness}

\begin{figure}
\plotone{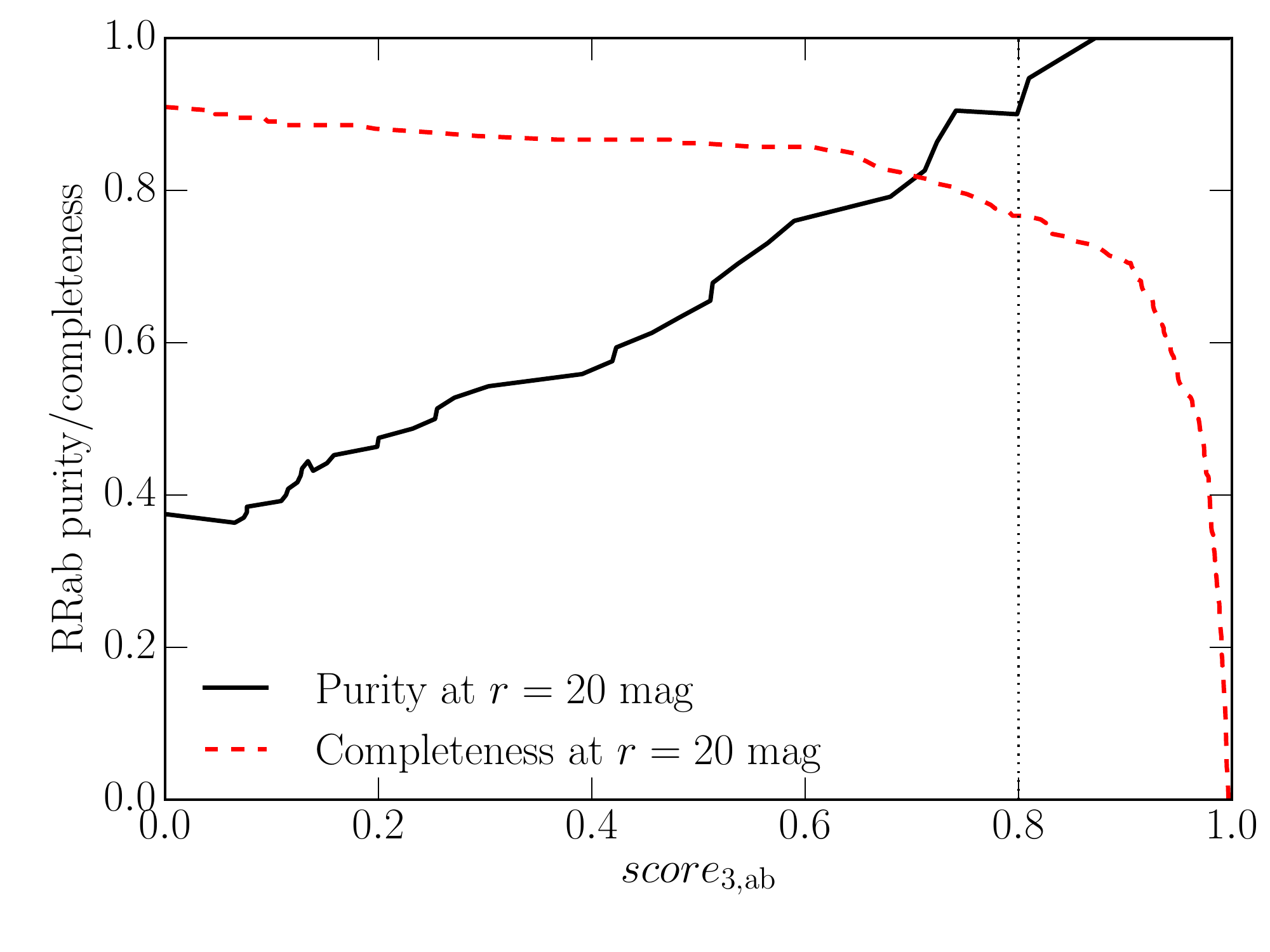}
\caption{
The expected purity and completeness for faint RRab stars, shown as a function of the threshold on the RRab classification score, $score_{\rm 3,ab}$. The initial purity is $38\%$ due to the $score_{\rm 2}>0.13$ requirement (\autoref{third_classifier}). A threshold of 0.8 (i.e., $score_{\rm 3,ab} > 0.8$, vertical dotted line) returns an RRab sample that is $91\%$ pure and $77\%$ complete at $\sim80$ kpc ($\langle r \rangle\sim20$ mag), 
\label{purity_completeness_faint}}
\end{figure}

The solid line in \autoref{purity_completeness_faint} shows the purity of the RRab selection at the faint end (at $\sim80$ kpc or $r\sim20$ mag), given a threshold on $score_{\rm 3, ab}$. To make this curve, we use 80 labeled objects from the SDSS Stripe 82 training set with $19.7 < \langle r \rangle< 20.3$ and $score_{\rm 2} > 0.13$, and calculate the fraction of true RRab stars in selected samples (given a threshold on $score_{\rm 3, ab}$). 

\capstartfalse
\begin{deluxetable}{ccc}
\tablecolumns{3}
\tablecaption{Expected RRab selection purity and completeness at $\sim80$ kpc ($\langle r \rangle\sim20$ mag)\label{table_RRab_faint}}
\tablehead{
\colhead{Threshold on $score_{\rm 3,ab}$} & \colhead{Purity} & \colhead{Completeness}
}
\startdata
0.00 & 0.38 & 0.91 \\
0.05 & 0.37 & 0.90 \\
0.10 & 0.39 & 0.89 \\
0.15 & 0.44 & 0.89 \\
0.20 & 0.47 & 0.88 \\
0.25 & 0.50 & 0.88 \\
0.30 & 0.54 & 0.87 \\
0.35 & 0.55 & 0.87 \\
0.40 & 0.56 & 0.87 \\
0.45 & 0.61 & 0.87 \\
0.50 & 0.65 & 0.86 \\
0.55 & 0.72 & 0.86 \\
0.60 & 0.76 & 0.86 \\
0.65 & 0.78 & 0.84 \\
0.70 & 0.81 & 0.82 \\
0.75 & 0.90 & 0.80 \\
0.80 & 0.91 & 0.77 \\
0.85 & 0.98 & 0.74 \\
0.90 & 1.00 & 0.71 \\
0.95 & 1.00 & 0.55
\enddata
\tablecomments{A machine readable version of this table with a 0.01 step in threshold, is available in the electronic edition of the Journal.} 
\end{deluxetable}
\capstarttrue

To quantify the completeness of the RRab selection at the faint end, we use 242 RRab stars from the Draco dSph training set (see \autoref{training_set}) that have $score_{\rm 2} > 0.13$. The dashed line in \autoref{purity_completeness_faint} shows the completeness of the selection (i.e., the fraction of recovered RRab stars) as a function of the threshold on $score_{\rm 3, ab}$. This completeness includes all losses due to initial data quality cuts, and classification cuts (i.e., $score_{\rm 1} > 0.01$ and $score_{\rm 2} > 0.13$). For convenience, we tabulate the purity and completeness in \autoref{table_RRab_faint}.

Above, we have used the Stripe 82 sample to measure the purity, and the Draco sample to measure the completeness. We did so because the S82 sample covers a large area and thus contains a more representative sample of contaminants that we may expect to encounter elsewhere on the sky. The Draco sample was used because it contains more faint ($r~\sim20$ mag) RR Lyrae stars than the Stripe 82 sample, and thus the estimate of the completeness has a lower Poisson noise.

Given the sparseness and the multi-band nature of PS1 data, it is remarkable that our selection method can deliver samples of RRab stars that are $\sim90\%$ pure and $\sim80\%$ complete (e.g., for $score_{\rm 3, ab} > 0.8$), even at distances as far as $\sim80$ kpc from the Sun. This raises hope for even better performance at the bright end.

\begin{figure}
\plotone{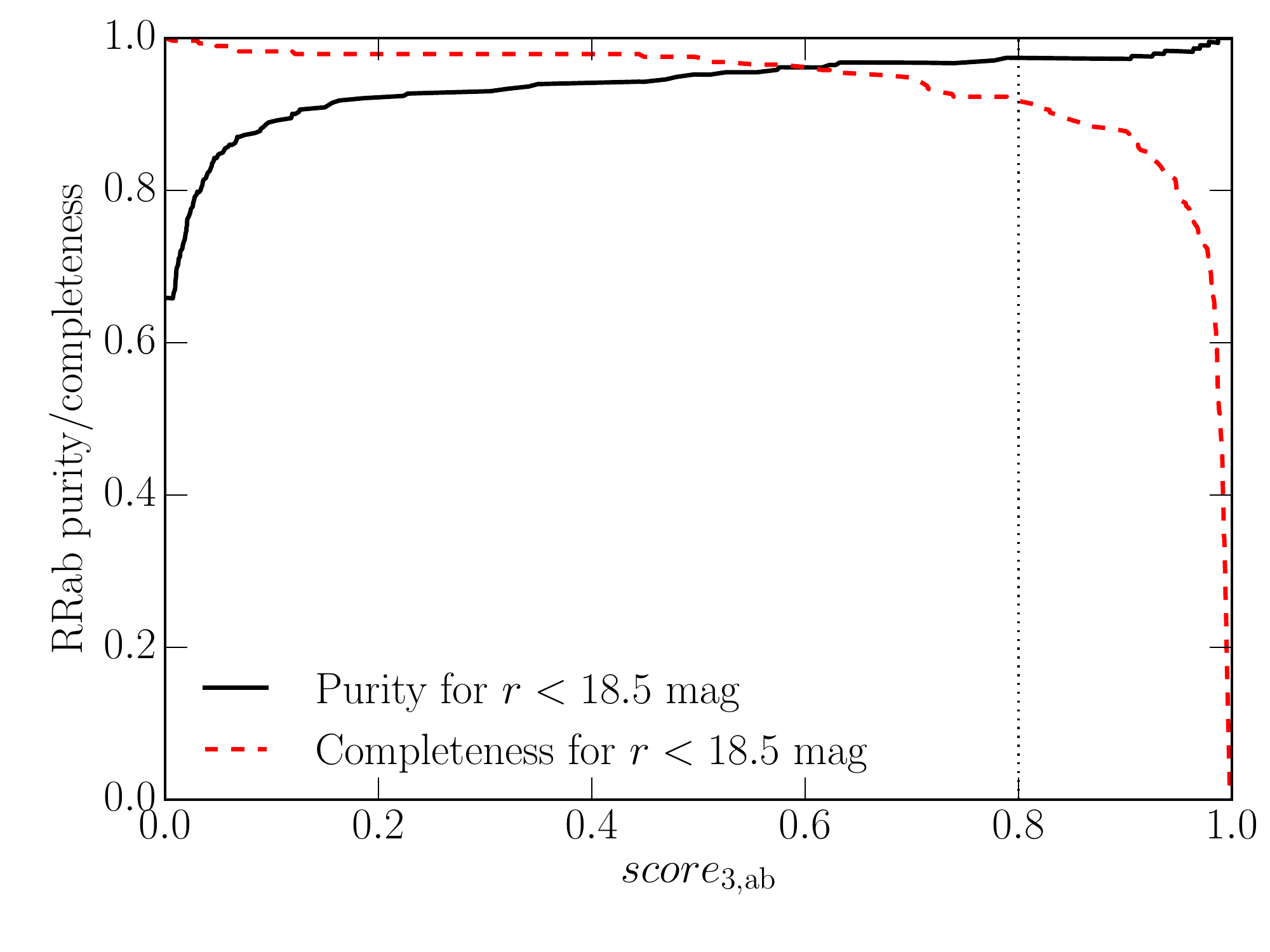}
\caption{
The expected purity and completeness of selected samples of bright RRab stars, as a function of the threshold on the RRab classification score, $score_{\rm 3,ab}$. A threshold of 0.8 (vertical dotted line) returns an RRab sample that is $97\%$ pure and $92\%$ complete within $\sim40$ kpc ($\langle r \rangle\lesssim18.5$ mag). 
\label{purity_completeness_bright_ab}}
\end{figure}

To measure the purity and completeness at the bright end, we select objects from the SDSS Stripe 82 training set with $score_{\rm 2}>0.13$ and $\langle r \rangle<18.5$ mag (i.e., within $\sim40$ kpc from the Sun). We use the $\langle r \rangle=18.5$ mag brightness cut because the vast majority of halo RR Lyrae stars are located within that magnitude range \citep{ses10}. The relevant curves are plotted in \autoref{purity_completeness_bright_ab} and tabulated in \autoref{table_RRab_bright}.

\capstartfalse
\begin{deluxetable}{ccc}
\tablecolumns{3}
\tablecaption{Expected RRab selection purity and completeness within $\sim40$ kpc ($\langle r \rangle<18.5$ mag)\label{table_RRab_bright}}
\tablehead{
\colhead{Threshold on $score_{\rm 3,ab}$} & \colhead{Purity} & \colhead{Completeness}
}
\startdata
0.00 & 0.66 & 1.00 \\
0.05 & 0.85 & 0.99 \\
0.10 & 0.89 & 0.98 \\
0.15 & 0.91 & 0.98 \\
0.20 & 0.92 & 0.98 \\
0.25 & 0.93 & 0.98 \\
0.30 & 0.93 & 0.98 \\
0.35 & 0.94 & 0.98 \\
0.40 & 0.94 & 0.98 \\
0.45 & 0.94 & 0.98 \\
0.50 & 0.95 & 0.97 \\
0.55 & 0.96 & 0.97 \\
0.60 & 0.96 & 0.96 \\
0.65 & 0.97 & 0.95 \\
0.70 & 0.97 & 0.95 \\
0.75 & 0.97 & 0.92 \\
0.80 & 0.97 & 0.92 \\
0.85 & 0.97 & 0.89 \\
0.90 & 0.97 & 0.88 \\
0.95 & 0.98 & 0.80
\enddata
\tablecomments{A machine readable version of this table with a 0.01 step in threshold, is available in the electronic edition of the Journal.} 
\end{deluxetable}
\capstarttrue

\begin{figure}
\plotone{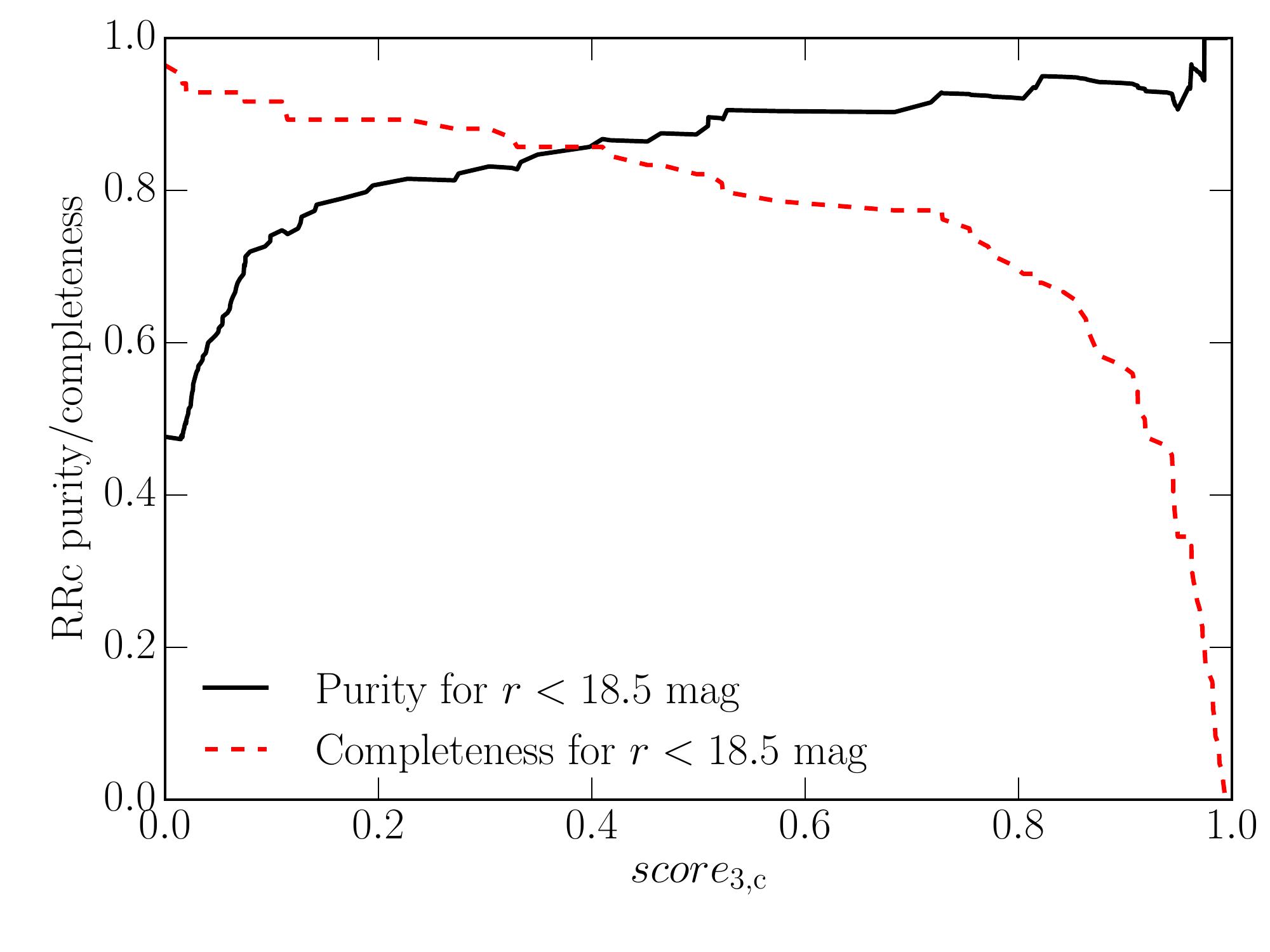}
\caption{
The expected purity and completeness of selected samples of bright RRc stars, as a function of the threshold on the RRc classification score, $score_{\rm 3,c}$. In comparison with RRab stars (\autoref{purity_completeness_bright_ab}), the selection of pure and complete samples of RRc stars is more challenging.
\label{purity_completeness_bright_c}}
\end{figure}

Finally, the purity and completeness curves characterizing the selection of bright RRc stars are shown in \autoref{purity_completeness_bright_c} and tabulated in \autoref{table_RRc_bright}. \autoref{purity_completeness_bright_c} shows that the selection of pure and complete samples of RRc stars is more challenging, both due to the lower amplitude of the pulsation and to the contamination by contact binaries with similar sinusoidal light curves. Nonetheless, it is still possible to produce samples that are over 80\% complete and pure within $\sim40$ kpc from the Sun. We do not discuss RRc stars further as they are less numerous than RRab stars (by a factor of three), and thus are of lesser importance for Galactic studies. 

\capstartfalse
\begin{deluxetable}{ccc}
\tablecolumns{3}
\tablecaption{Expected RRc selection purity and completeness within $\sim40$ kpc ($\langle r \rangle<18.5$ mag)\label{table_RRc_bright}}
\tablehead{
\colhead{Threshold on $score_{\rm 3,c}$} & \colhead{Purity} & \colhead{Completeness}
}
\startdata
0.00 & 0.48 & 0.96 \\
0.05 & 0.61 & 0.93 \\
0.10 & 0.74 & 0.92 \\
0.15 & 0.78 & 0.89 \\
0.20 & 0.81 & 0.89 \\
0.25 & 0.81 & 0.89 \\
0.30 & 0.83 & 0.88 \\
0.35 & 0.85 & 0.86 \\
0.40 & 0.86 & 0.86 \\
0.45 & 0.86 & 0.83 \\
0.50 & 0.88 & 0.82 \\
0.55 & 0.90 & 0.79 \\
0.60 & 0.90 & 0.78 \\
0.65 & 0.90 & 0.78 \\
0.70 & 0.91 & 0.77 \\
0.75 & 0.93 & 0.75 \\
0.80 & 0.92 & 0.70 \\
0.85 & 0.95 & 0.66 \\
0.90 & 0.94 & 0.57 \\
0.95 & 0.91 & 0.35
\enddata
\tablecomments{A machine readable version of this table with a 0.01 step in threshold, is available in the electronic edition of the Journal.} 
\end{deluxetable}
\capstarttrue

\subsection{RRab Selection Function}\label{selection_function}

Given a position on the sky and the flux-averaged $r$-band magnitude of an RR Lyrae star, what is the probability of selecting that star using the PS1 data at hand? Characterizing this selection function is of obvious importance for studies of the Galactic structure, especially when modeling the number density distribution of stars (e.g., \citealt{bov12, xue15}). In this Section we restrict ourselves to characterizing the selection function of RRab stars at high galactic latitudes ($|b| > 20\arcdeg$), because i) they are three times more numerous than RRc stars, and ii) at a given purity, they can be recovered at a much higher rate than RRc stars (compare \autoref{purity_completeness_bright_ab} vs.~\autoref{purity_completeness_bright_c}). Characterizing the selection function at low galactic latitudes would require an appropriate training set (Stripe 82 and Draco are both located at $|b| > 20\arcdeg$).

\begin{figure}
\plotone{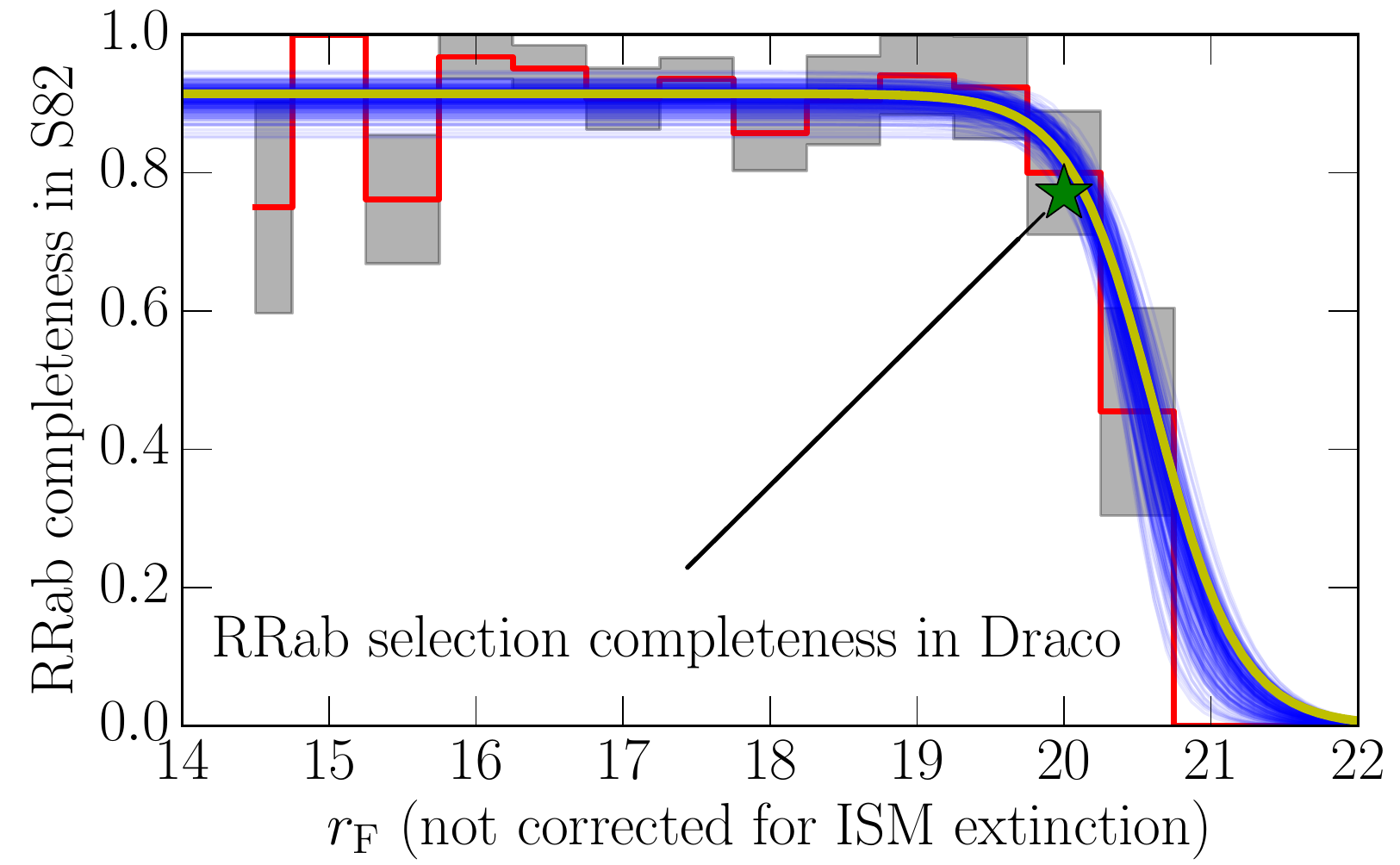}
\caption{
The RRab selection function, or the completeness of the selection of RRab stars at high galactic latitudes ($|b| >20\arcdeg$), as a function of the flux-averaged $r$-band magnitude (not corrected for interstellar matter extinction). The red curve shows the ratio of the number of selected ($n_{\rm sel}$, $score_{\rm 3,ab}>0.8$) and all RRab stars from the SDSS Stripe 82 training set ($n_{\rm all}$), in 0.5 mag wide bins. The shaded region shows the standard deviation of the bin height, $\sqrt{n_{\rm sel}(1 - n_{\rm sel}/n_{\rm all})}/n_{\rm all}$, computed based on the binomial distribution. For comparison, the star symbol shows the fraction of recovered RRab stars in the Draco dSph (see \autoref{purity_completeness_faint}). The thick yellow line shows the best-fit logistic curve (\autoref{logistic_curve}, $L = 0.91$, $k = 4.1$, $x_{\rm 0} = 20.57$ mag), and the thin blue lines illustrate the uncertainty of the fit (see \autoref{selection_function} for details).
\label{completeness_rmag}}
\end{figure}

We assume that the selection function $S$ depends only on the flux-averaged $r$-band magnitude of an RRab star (not corrected for interstellar extinction), and not on its position (i.e., $S(r_{\rm F}$)). This is a reasonable assumption given the uniformity of dust extinction away from the Galactic plane, and the uniformity of PS1 multi-epoch coverage. The selection function will also depend on the threshold imposed on the final classification score, $score_{\rm 3, ab}$. For the sake of simplicity we only consider the case when $score_{\rm 3, ab} > 0.8$, as this selection cut returns a sample that is appropriate for many studies (90\% purity and 80\% completeness, even at the faint end; see \autoref{section_purity_completeness}). By assuming spatial independence, we can now use the SDSS Stripe 82 and Draco training sets to determine the PS1 $3\pi$ selection function of RRab stars at high galactic latitudes. The result is illustrated in \autoref{completeness_rmag}.

We find that the RRab selection function is approximately constant at $\sim90\%$ for $r_{\rm F}\lesssim20$ mag, after which it steeply drops to zero at $r_{\rm F}\sim21.5$ mag. To characterize the selection function, we construct a simple probabilistic model.

There are 577 RRab stars in our training set (in SDSS Stripe 82 and Draco), of which 483 pass the $score_{\rm 3,ab}>0.8$ selection cut. With each RRab star we associate a $(r_{\rm F, n}, s_{\rm n})$ pair of values, where $s_{\rm n} = 1$ if the star is selected, otherwise $s_{\rm n} = 0$ ($r_{\rm F,n}$ is the star's extincted flux-averaged $r$-band magnitude). We denote the full data set of 577 pairs of values as ${\bf d_{\rm n}} = \{ r_{\rm F,n}, s_{\rm n} \}$.

The likelihood of this data set is given by
\begin{equation}
p\left({\bf d_{\rm n}} | L, k, x_{\rm 0}\right) = \prod_{\rm n = 1}^{577}p\left(s_{\rm n} | S\left(r_{\rm F,n} | L, k, x_{\rm 0}\right)\right),
\end{equation}
where $p\left(s_{\rm n} | S\left(r_{\rm F,n} | L, k, x_{\rm 0}\right)\right)$ is the Bernoulli probability mass function with success probability given by the selection function, $S\left(r_{\rm F,n} | L, k, x_{\rm 0}\right)$.

To model the selection function, we use the logistic curve
\begin{equation}
S(r_{\rm F, n} | L, k, x_{\rm 0}) = \frac{L}{1 + \exp(-k(r_{\rm F, n} - x_{\rm 0}))},\label{logistic_curve}
\end{equation}\\
where $L$ is the curve's maximum value, $k$ is the steepness of the curve, and $x_{\rm 0}$ is the magnitude at which the completeness drops to $50\%$.

The probability of this model given data ${\bf d_{\rm n}}$ is then
\begin{equation}
p(L, k, x_{\rm 0} | {\bf d_{\rm n}}) = p({\bf d_{\rm n}} | L, k, x_{\rm 0})p(L, k, x_{\rm 0}),
\end{equation}
where $p(L, k, x_{\rm 0})$ is the prior probability of model parameters. We impose uniform priors such that $0.4 \leq S(r_{\rm F} < 18.5) \leq 1.0$ (i.e., completeness within 40 kpc is between 40\% and 100\%) and $S(r_{\rm F} > 22) = 0$.

We explore the probability of various model parameters using the \citet{gw10} Affine Invariant Markov chain Monte Carlo (MCMC) ensemble sampler, as implemented in the \texttt{emcee} package\footnote{\url{http://dan.iel.fm/emcee/current/}} (v2.2.1, \citealt{fm13}). The most probable model of the selection function (yellow curve) is shown in \autoref{completeness_rmag}, with the best-fit logistic curve (\autoref{logistic_curve}) being $L = 0.91$, $k = 4.1$, $x_{\rm 0} = 20.57$ mag. To illustrate the uncertainty in the model, we also plot the curves associated with 200 randomly selected models from the posterior distribution (thin blue lines).

\section{PS1 Catalog of RR Lyrae Stars}\label{catalog}

We have applied the above multi-step selection procedure to about 500 million PS1 objects that pass PS1 data quality cuts (\autoref{Sec:PS1_3pi}), and have calculated final RRab and RRc classification scores ($score_{\rm 3,ab}$, and $score_{\rm 3,c}$) for 240,000 objects. We report their positions, distances, PS1 photometry, and classification scores in \autoref{PS1_RRLyrae_table}. A total of $\sim400,000$ CPU hours of super-computing time was used to process all of the data and calculate the final classification scores. Below we illustrate some properties of this sample and leave a more detailed analysis of the distribution of RR Lyrae stars in the Galactic halo for future studies.

\begin{figure*}
\plotone{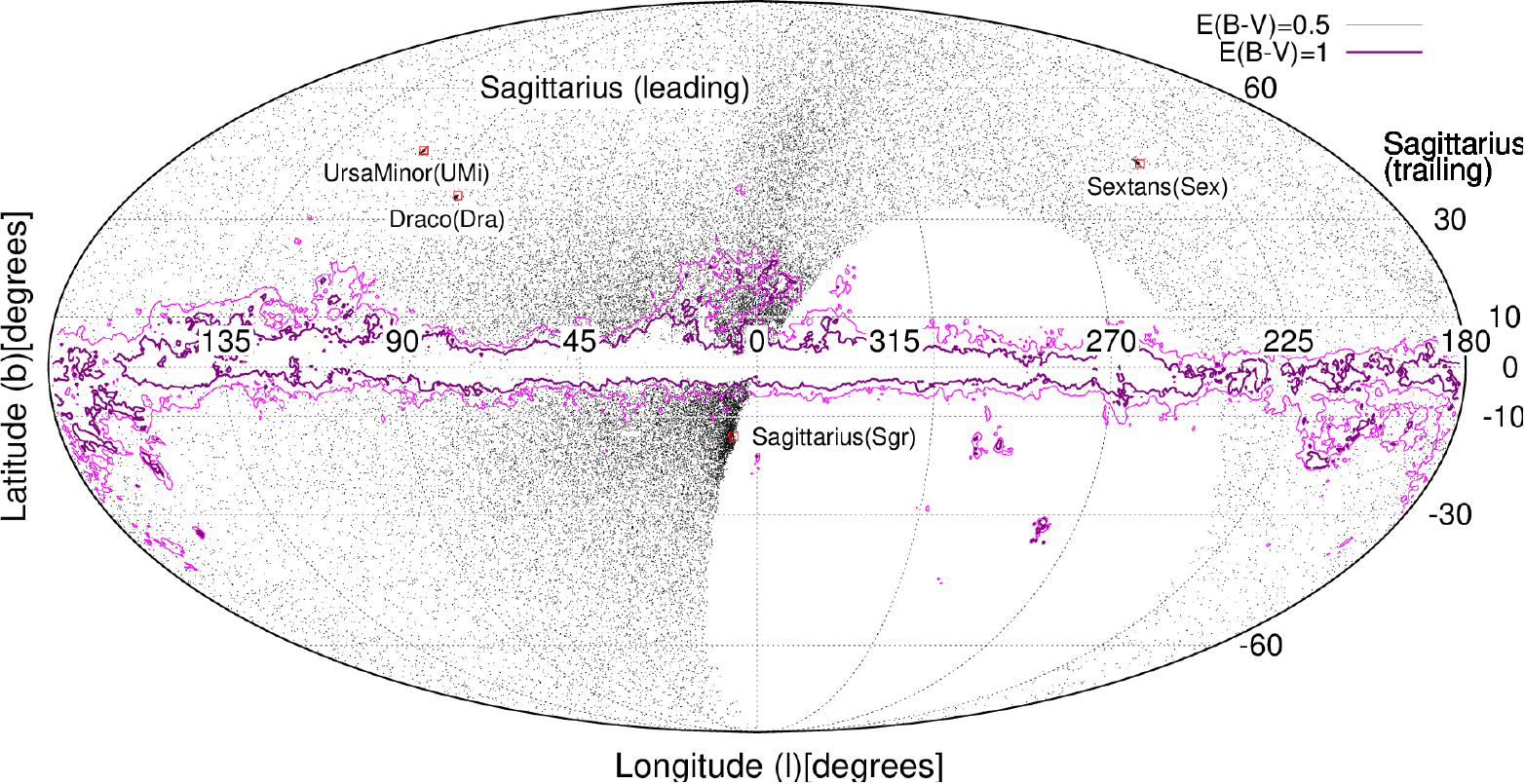}
\caption{
Distribution of $\sim45,000$ highly probable RRab stars ($score_{\rm 3,ab} > 0.8$, expected purity of 90\% and completeness of $\sim80\%$ at 80 kpc), shown in Mollweide projection of Galactic coordinates. A contour plot of the reddening-based $E(B-V)$ dust map \citep{sch14} is overlayed, as well as the positions of four Milky Way dwarf satellite galaxies. The locations of the leading and trailing arms of the Sagittarius tidal stream are also indicated.
\label{mollweide_projection}}
\end{figure*}

To illustrate the coverage of the PS1 catalog of RR Lyrae stars, we have selected a sample of $\sim 45,000$ highly probable RRab stars ($score_{3,ab} > 0.8$, expected purity of 90\% and completeness of $\sim80\%$ at 80 kpc), and have plotted their angular distribution in \autoref{mollweide_projection}.

The leading arm of the Sagittarius tidal stream \citep{iba01} and four Milky Way satellite galaxies are most easily discernible features in \autoref{mollweide_projection}. However, another notable feature is an almost complete absence of {\em high probability} RRab stars (i.e., those that have $score_{\rm 3,ab} > 0.8$) in regions with high ISM extinction (e.g., $E(B-V) > 1$). Improperly dereddened photometry is the most likely reason for this lack of high probability RRab stars at low galactic latitudes.

Briefly, when dereddening photometry, we assume that all sources are located behind the dust layer. At low galactic latitudes this may not always be true, as sources may be embedded in the dust layer. After dereddening, the photometry of such sources will be overcorrected for extinction, and their optical PS1 colors will be shifted blueward from their dust-free values. In addition, improperly dereddened light curves will not be well-fit by multiband templates (\autoref{multi-band_light_curve_fitting}). As a result of these effects, true RR Lyrae stars may look like non-RR Lyrae objects, and may end up with low $score_{\rm 3, ab}$ values.

The lack of high probability RRab stars at low galactic latitudes also demonstrates the resilience of the classifier to contamination. Due to the increase in stellar number density, and the fact that some fraction of star will be incorrectly tagged as RR Lyrae stars, one would naively expect for the density of objects tagged as RR Lyrae stars to increase towards the Galactic plane. However, no such increase is observed in \autoref{mollweide_projection}. The features extracted during multiband template fitting are most likely responsible for this resilience, as even a significant increase in the number of contaminants is not sufficient to produce objects that match multiband light curve characteristics of RR Lyrae stars.

\begin{figure}
\plotone{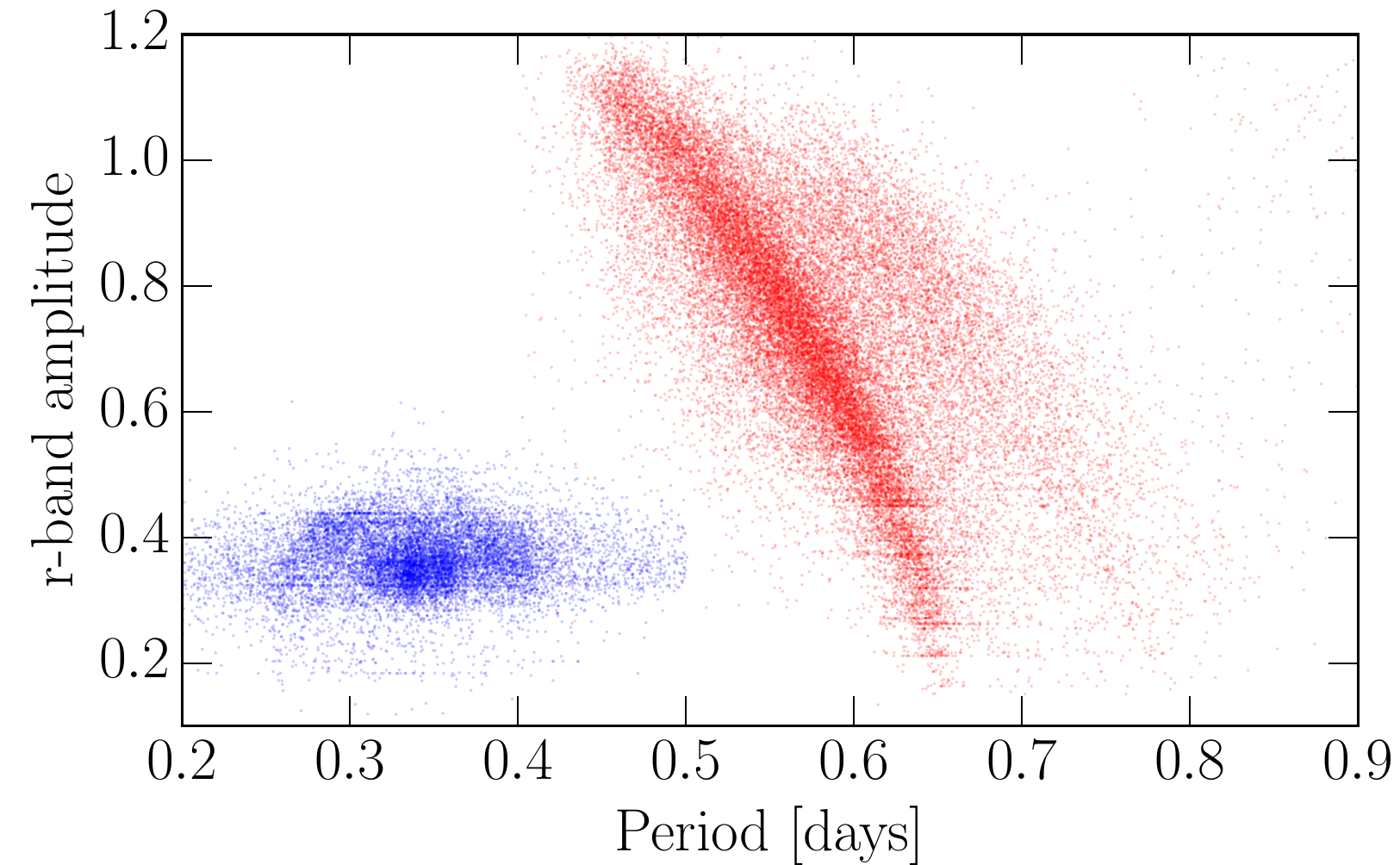}
\caption{
The distribution of highly likely RRab (red; $score_{\rm 3, ab} > 0.8$) and RRc stars (blue; $score_{\rm 3, c} > 0.55$) in the period vs.~$r_{\rm P1}$-band amplitude diagram. Note the well-defined Oosterhoff I locus of RRab stars \citep{oos39,cat09}, and the less populated Oosterhof II locus shifted to longer periods (along the lines of constant amplitude). The apparent clumps of RRc stars are likely caused by period aliasing (e.g., the 1-day beat frequency alias, see the top panel of \autoref{period_comparison}).
\label{period_amplitude}}
\end{figure}

And finally, to illustrate the efficiency of the final multiclass classifier (\autoref{third_classifier}) at separating RRab and RRc stars, we show their distribution in the period vs.~$r_{\rm P1}$-band amplitude diagram (\autoref{period_amplitude}).

\section{Discussion and Summary}

In this paper, we have explored with what fidelity RR Lyrae stars can be identified in multi-epoch, asynchronous multi-band photometric data. We have done this for the specific case of the PS1 $3\pi$ light curves which are very sparse; $\lesssim12$ epochs per band over 3000 typical RR Lyrae pulsation periods (i.e., 4 years). To identify RR Lyrae stars, we have employed, in particular, the fitting (and period phasing) of very specific empirical RR Lyrae light curves, and have utilized supervised machine-learning tools. While we have applied our selection only to this specific data set, many of the approaches described here will be applicable to other sparse, asynchronous multi-band data sets, such as those produced by the Dark Energy Survey (DES; \citealt{DES}) and the Large Synoptic Survey Telescope (LSST; \citealt{ive08}). For example, for its Galactic plane sub-survey, LSST is currently planning to obtain 12 observations over 4 years in each of its $ugrizy$ bandpasses (i.e., 30 observations per band over 10 years; \v{Z}.~Ivezi\'{c}, priv.~comm.), making its data set very similar to the one produced by the PS1 $3\pi$ survey. Compared to PS1, however, LSST will go deeper by at least 2 mag in $izy$ bands, allowing us to select RR Lyrae stars and study old stellar populations close to the Galactic plane, even at the far side of the Galaxy.

We demonstrated that we can precisely and accurately measure the periods (to within 2 seconds) for the vast majority of RR Lyrae within PS1 $3\pi$, extending to distance moduli of $\sim20$, or $\sim 100$~kpc. The high precision of the period determination may seem surprising at face-value, but is owed to the long time-baseline: a 2s/period difference causes a cumulative light curve shift of 90 minutes across 3000 pulsation periods. Accurate periods are crucial for calculating the phase of spectroscopic observations, and for transforming the observed radial velocity to the center-of-mass velocity needed for kinematic studies \citep{sesar12}. The ephemerides (i.e., periods and phase offsets) provided by our catalog can thus be readily used to turn RR Lyrae stars observed by current (e.g., SDSS-IV/TDSS; \citealt{rua16}) and upcoming multi-object spectroscopic (MOS) surveys (e.g., Gaia, WEAVE, DESI; \citealt{per01,dal14,lev13}) into precise kinematic tracers of the halo structure and substructure (i.e., stellar streams). With a density of 1 deg$^{-1}$, PS1 RR Lyrae stars represent a unique ``piggyback'' project for MOS surveys, with a potentially high impact and certainly low cost ($\sim1$ target per MOS field).

Using these light curve fits as one (crucial) feature in a supervised classification of RR Lyrae, we showed that we can -- at least at high Galactic latitudes -- construct a sample of $\sim45,000$ RRab stars that has 90\% purity and 80\% completeness, even at 80 kpc from the Sun. In comparison with previous catalogs, our sample is deeper than the SDSS Stripe 82 sample of \citet{ses10}, while covering more of the sky than the CRTS sample of \citet{dra13}. The PS1 $3\pi$ data and the classification presented here even allow for a quite reliable separation of RRab and RRc type of RR Lyrae stars, as shown by the period-amplitude diagram (\autoref{period_amplitude})

All this opens up many avenues in exploring the Galactic halo. With its second data release (DR2) expected in April 2018, Gaia astrometric mission will provide unprecedented proper motions for PS1 RR Lyrae stars brighter than $V\sim20$ mag, but no competitive distance information (beyond a few kpc). Having precise distances is crucial for measuring tangential velocities\footnote{Radial velocities of RR Lyrae stars are straightforward to measure \citep{sesar12}.}, and thus the Galactic potential, as the uncertainty in tangential velocity increases proportionally with the uncertainty in distance. Therefore, it is particularly remarkable that, using PS1 data and a period-absolute magnitude relation, we can measure distances to RR Lyrae stars with a precision of $\sim3\%$, even for stars at 100 kpc from the Sun.

An important avenue to explore with the resulting RR Lyrae catalog is the question of RR Lyrae at low-latitudes (covered by PS1). These objects open up the possibility to explore the oldest portion of the Galactic disk. At low latitudes, the selection function of the sample will, however, be considerably more complicated, warranting careful testing and characterization beyond the scope of this paper.

\capstartfalse
\begin{deluxetable*}{ccccccccccccc}
\tabletypesize{\footnotesize}
\tablecolumns{13}
\tablecaption{PS1 Catalog of RR Lyrae Stars\label{PS1_RRLyrae_table}}
\tablehead{
\colhead{R.A.} & \colhead{Decl.} & \colhead{$score_{\rm 3,ab}^a$} & \colhead{$score_{\rm 3,c}^a$} & \colhead{DM$^b$} & \colhead{Period} & \colhead{$\phi_0^c$} & \colhead{$A^{\prime}_{g}$ \dots $A_{z}^{\prime,d}$} & \colhead{$g^{\prime}$ \dots $z^{\prime,e}$} & \colhead{$T_g$ \dots $T_z^f$} & \colhead{$g_F$ \dots $z_F^g$} & \colhead{E(B-V)$^h$}
}
\startdata
180.39736 & -0.23480 & 0.57 & 0.02 & 15.44 & 0.671302 & -0.40301 & 0.22 \dots 0.11 & 15.96 \dots 15.77 & 100 \dots 100 & 16.09 \dots 15.82 & 0.020 \\
179.98457 & -0.00105 & 0.99 & 0.00 & 15.90 & 0.471807 & -0.18385 & 1.32 \dots 0.68 & 15.72 \dots 16.12 & 120 \dots 113 & 16.75 \dots 16.58 & 0.012
\enddata
\tablenotetext{a}{Final RRab and RRc classification scores.}
\tablenotetext{b}{Distance modulus calculated using the flux-averaged $i_{\rm P1}$-band magnitude and \autoref{abs_mag_i_band}. The uncertainty in distance modulus is $0.06(rnd)\pm0.03(sys)$ mag {\em for RRab stars}. This distance modulus may be biased and more uncertain for RRc stars.}
\tablenotetext{c}{Phase offset (see \autoref{phase}).}
\tablenotetext{d}{Best-fit amplitude (e.g., $A_g^\prime = FA_g$; see \autoref{multi-band_templates_equations}).}
\tablenotetext{e}{Best-fit magnitude at $\phi=0$, {\em corrected} for dust extinction using extinction coefficients of \citet{sf11} and the dust map of \citet{sch14} (e.g., $g^\prime = g_0 - r_0 + r^\prime$); see \autoref{multi-band_templates_equations}).}
\tablenotetext{f}{Best-fit template ID number (see Section 3.1 and Table 2 of \citealt{ses10}).}
\tablenotetext{g}{Flux-averaged magnitude, {\em corrected} for dust extinction using extinction coefficients of \citet{sf11} and the dust map of \citet{sch14}.}
\tablenotetext{h}{Reddening adopted from the \citet{sch14} dust map.}
\tablecomments{A machine readable version of this table will become available on Nov 1 2017 in the electronic edition of the Journal. A portion is shown here for guidance regarding its form and content. For collaborations on projects and earlier access to the PS1 catalog of RR Lyrae stars, please contact the first author.} 
\end{deluxetable*}
\capstarttrue

\acknowledgments

B.S., N.H. and H.-W.R.~acknowledge funding from the European Research Council under the European Union’s Seventh Framework Programme (FP 7) ERC Grant Agreement n.~${\rm [321035]}$. H.-W.R. acknowledges support of the Miller Institute at UC Berkeley through a visiting professorship during the completion of this work. We thank the anonymous referee for the thorough review, positive comments, and constructive remarks on this manuscript. The Pan-STARRS1 Surveys (PS1) have been made possible through contributions by the Institute for Astronomy, the University of Hawaii, the Pan-STARRS Project Office, the Max-Planck Society and its participating institutes, the Max Planck Institute for Astronomy, Heidelberg and the Max Planck Institute for Extraterrestrial Physics, Garching, The Johns Hopkins University, Durham University, the University of Edinburgh, the Queen's University Belfast, the Harvard-Smithsonian Center for Astrophysics, the Las Cumbres Observatory Global Telescope Network Incorporated, the National Central University of Taiwan, the Space Telescope Science Institute, and the National Aeronautics and Space Administration under Grant No.~NNX08AR22G issued through the Planetary Science Division of the NASA Science Mission Directorate, the National Science Foundation Grant No.~AST-1238877, the University of Maryland, Eotvos Lorand University (ELTE), and the Los Alamos National Laboratory. 

\appendix

\section{Features Extracted From Phased Multi-band Light Curves}\label{extraction_of_features}

In this Section, we describe features extracted from phased multi-band light curves.

\begin{itemize}
\item {\em rrab}: This flag has a value of 1 if the best-fit template is of type $ab$, and 0 otherwise.
\item {\em period}: the best-fit period associated with the best-fit template (i.e., the multi-band template with the smallest $\chi^2$ value).
\item $\chi^2$ and $\chi_{\rm dof}^2$: goodness-of-fit $\chi^2$ and $\chi^2$ per degree of freedom associated with the best-fit template.
\item $L_{\rm max}$ and $L_{\rm max, dof}$: To calculate these two values, we sort $\chi^2$ and $\chi_{\rm dof}^2$ values (measured during the first fitting run) into 1 sec-wide bins in period (e.g., see \autoref{template_fit}), and for each bin calculate $L(bin) = \sum \exp(-\chi^2/2)$ and $L_{\rm dof}(bin)=\sum \exp(-\chi_{\rm dof}^2/2)$, where the sum goes over all $\chi^2$ and $\chi_{\rm dof}^2$ in a bin. Then, $L_{\rm max} = \max(L(bin))$ and $L_{\rm max, dof} = \max(L_{\rm dof}(bin))$.
\item Stetson $J$ index: We phase a PS1 multi-band light curve using its best-fit period, and calculate normalized residuals $\delta(phase) = (m_{\rm n} - m(phase(t_{\rm n}, period, \phi))/\sigma_{\rm m_{\rm n}}$, where $m_{\rm n}$ and $\sigma_{\rm m_{\rm n}}$ are the magnitude and uncertainty of the $n$-th observation in the $m=g,r,i,z$  band, and $m(phase(t_{\rm n}, period, \phi))$ is the magnitude predicted by the best-fit model at the phase of the $n$-th observation. The $\delta$ values are then sorted by phase, and used to calculate the Stetson $J$ index (see Section 2 of \citealt{ste96}).
\item Entropy of the phase light curve $S_{\rm phased}$: We phase a PS1 multi-band light curve using its best-fit period, bin observations into 20 phase bins (ignoring the band), and calculate $S_{\rm phased}= -\sum_{bin} (N_{\rm bin}/N_{\rm tot})\ln(N_{\rm bin}/N_{\rm tot})$, where $N_{\rm bin}$ is the number of observations in a phase bin, and $N_{\rm tot}$ is the total number of observations in the light curve.
\item $\Delta P$: The difference between the best-fit period and the period associated with the best-fit multi-band template (i.e., the corresponding RR Lyrae star in SDSS Stripe 82; listed in Table 2 of \citealt{ses10}).
\item $\Delta(g-r)_0$, $\Delta(r-i)_0$, $\Delta(r-z)_0$: The difference between the color measured from the best-fit {\em model} at $phase=0$, and the color measured from the best-fit {\em template}. Since the fitting parameter $F$ allows the amplitudes in each band to vary by up to 20\% with respect to amplitudes associated with the best-fit multi-band template, the above differences do not have to be zero.
\item $(g-r)_0$, $(r-i)_0$, $(r-z)_0$: Color at $phase=0$, measured from the best-fit model.
\item $FA_{\rm g}$, $FA_{\rm r}$, $FA_{\rm i}$, $FA_{\rm z}$: best-fit amplitudes of $griz$ bands, where $A_{\rm m}$ is the amplitude associated with the $m=g, r, i, z$ band of the best-fit multi-band template, and $F$ is a fitting parameter.
\end{itemize}

\section{Constraining Period-Absolute Magnitude-Metallicity Relations for PS1 Bands}\label{PLZ_derivation}

To constrain \autoref{PLZ}, we use average (apparent) PS1 $griz$ magnitudes of 55 RR Lyrae stars located in 5 Galactic globular clusters that have metallicities ranging from -1.02 dex to -2.37 dex (on the \citealt{car09} scale and taken from the \citealt{har96,har10} catalog; NGCs 6171, 5904, 4590, 6341, and 7078). The distance moduli of these globular clusters were measured by \citet[see their Table 4]{sol06} using a period-luminosity-metallicity (PLZ) relation whose zero point of $-1.05\pm0.13$ mag was constrained using the Hubble Space Telescope (HST) parallax of the star RR Lyrae. Using HST parallaxes of five RR Lyrae stars, \citet{ben11} measured the zero point of the \citet{sol06} PLZ relation to be $-1.08\pm0.03$ mag. Thus, for distance moduli of 55 RR Lyrae stars, we adopt the distance moduli of their globular clusters (see Table 4 of \citealt{sol06}), shifted by 0.03 mag to match the more precise zero point measured by \citet{ben11}. As the systematic uncertainty in distance moduli of calibration RR Lyrae stars, we assume a value of 0.03 mag.

We measure average (apparent) $griz$ magnitudes of RR Lyrae stars by first converting their best-fit $griz$ model light curves (e.g., see sold lines in \autoref{template_fit}) from magnitude to flux units. The model curves in flux units are then integrated over a pulsation cycle, and the integrated fluxes are expressed in units of magnitude. Hereafter, we call these {\em flux-averaged} magnitudes ($g_{\rm F}$, $r_{\rm F}$, etc.), and distinguish them from uncertainty-weighted average magnitudes ($\langle g \rangle$, $\langle r \rangle$, etc.).

Given the above data, we wish to calculate the probability $p(\theta^b | \mathcal{D}^b)$ of some set of model parameters $\theta^b = \{\alpha^b, \beta^b, M^b_{\rm ref}, \sigma^b_{\rm M}\}$ for the $b=g, r, i, z$ PS1 band, where $\mathcal{D}^b=\{{\bf d^b_k}\}$ is the full data set of 55 stars, and ${\bf d^b_k} = \{DM, {\rm [Fe/H]}, \log_{\rm 10}(P), b_{\rm F}\}$ is the data set associated with the $k$-th star that contains its distance modulus and metallicity (assumed to be the same as that of the cluster), period, and flux-averaged PS1 apparent magnitudes.

Using the relations of conditional probability, the probability $p(\theta^b | \mathcal{D}^b)$ can be written as
\begin{equation}
p(\theta^b | \mathcal{D}^b) \propto p(\mathcal{D}^b | \theta^b)p(\theta^b),
\end{equation}
where $p(\theta^b)$ is the prior probability of model parameters, and
\begin{equation}
p(\mathcal{D}^b | \theta^b) = \prod_k p({\bf d^b_k} | \theta^b)
\end{equation}
is the likelihood of the full data set given some values of model parameters, $\theta^b$.

To make the fitting of PLZ relations robust to possible outliers (e.g., due to an incorrectly measured period, which may happen for 13\% of RRab stars; \autoref{multi-band_light_curve_fitting}), we model the likelihood of the $k$-th data set using a mixture model (see Section 3 and Equation 17 of \citealt{hbl10})
\begin{equation}
p({\bf d^b_k} | \theta^b) = (1 - A^b_{\rm out})p_{\rm in}({\bf d^b_k} | \theta^b_{\rm in}) + A^b_{\rm out}p_{\rm out}({\bf d^b_k} | \theta^b_{\rm out}),
\end{equation}
where $p_{\rm in}$ and $p_{\rm out}$ are the likelihoods of drawing the data set ${\bf d^b_k}$ from the inlier or outlier distribution, respectively, and $A^b_{\rm out}$ is the mixing proportion.

The likelihood of drawing the data set ${\bf d^b_k}$ from the inlier distribution is
\begin{equation}
p_{\rm in}({\bf d^b_k} | \theta^b_{\rm in}) = \mathcal{N}(b_{\rm F} - DM | M^b, \sigma^b_{\rm M}),
\end{equation}
where $\theta^b_{\rm in} = \{\alpha^b, \beta^b, M^b_{\rm ref}, \sigma^b_{\rm M}\}$, $b_{\rm F}$ is the apparent flux-averaged magnitude in the $b=g, r, i, z$ PS1 band, $M^b$ is the absolute magnitude in the $b$ band
\begin{equation}
M^b = \alpha^b\log_{\rm 10}(P/P_{\rm ref}) + \beta^b({\rm [Fe/H]} - {\rm [Fe/H]_{\rm ref}}) + M^b_{\rm ref},
\end{equation}

and
\begin{equation}
N(x | \mu, \sigma) = \frac{1}{\sqrt{2\pi\sigma^2}}e^{-\frac{(x-\mu)^2}{2\sigma^2}}.
\end{equation}

The likelihood of drawing the data set ${\bf d^b_k}$ from the outlier distribution is defined as
\begin{equation}
p_{\rm out}({\bf d^b_k} | \theta^b_{\rm out}) = \mathcal{N}(b_{\rm F} - DM | M^b_{\rm out}, \sigma^b_{\rm out}),
\end{equation}
where $M^b_{\rm out}$ has the same form as \autoref{PLZ}, but different parameters, $\theta^b_{\rm out} = \{\alpha^b_{\rm out}, \beta^b_{\rm out}, M^b_{\rm ref, out}, \sigma^b_{\rm out}\}$. Most importantly, the distribution of outliers is assumed to be wider than the distribution of inliers, that is, $\sigma^b_{\rm out} > \sigma^b_{\rm M}$.

Before we can calculate probability $p(\theta^b | \mathcal{D}^b)$, we need to define the prior probabilities of model parameters, $p(\theta^b)$. For the $\alpha^b$ parameter and $r_{\rm P1}$ and $i_{\rm P1}$ bands, we adopt informative Gaussian priors based on the slopes of PLZ relations in the globular cluster M4 \citep{bra15}: $p(\alpha^r | r_{\rm P1}) = \mathcal{N}(\alpha^r | -1.5, 0.1)$ and $p(\alpha^i | i_{\rm P1}) = \mathcal{N}(\alpha^i | -1.72, 0.07)$. For $\sigma^b_{\rm M}$ and $\sigma^b_{\rm out}$, we adopt Jeffreys log-uniform priors \citep{jay68}, and require that $\sigma^b_{\rm out} > \sigma^b_{\rm M}$. Uniform priors are adopted for the remaining model parameters and bands.

To efficiently explore the parameter space, we use the \citet{gw10} Affine Invariant Markov chain Monte Carlo (MCMC) Ensemble sampler as implemented in the \texttt{emcee} package\footnote{\url{http://dan.iel.fm/emcee/current/}} (v2.2.1, \citealt{fm13}). We use 200 walkers and obtain convergence\footnote{We checked for convergence of chains by examining the auto-correlation time of the chains per dimension.} after a short burn-in phase of 300 steps per walker. The chains are then evolved for another 2000 steps, and the first 300 (burn-in) steps are discarded.

To describe the marginal posterior distributions of individual model parameters, we measure the median, the difference between the 84th percentile and the median, and the difference between the median and the 16th percentile of each marginal posterior distribution (for a Gaussian distribution, these differences are equal to $\pm1$ standard deviation). We report these values in \autoref{PS1_PLZ} (\autoref{distance_precision}). We do not characterize parameters of the outlier distribution, and instead simply marginalize over them (i.e., we consider them as nuisance parameters).


\bibliography{main}

\end{document}